\documentclass[prx,twocolumn,floatfix]{revtex4-2}
\usepackage{bbold}
\usepackage{amssymb}
\usepackage{amsmath}
\usepackage{empheq}
\usepackage{amsthm}
\usepackage{siunitx}
\usepackage[normalem]{ulem}
\usepackage{caption}
\usepackage{subfig}
\usepackage[]{graphicx}

\usepackage{graphicx}% Include figure files
\usepackage{dcolumn}% Align table columns on decimal point
\usepackage{bm}% bold math
\graphicspath{ {Figures/} }

\usepackage{color}
\usepackage{float}
\begin{document}

\begin{large}
\noindent This is the accepted version of \textit{Jacopo M. De Ponti, Luca Iorio, Emanuele Riva, Raffaele Ardito, Francesco Braghin, and Alberto Corigliano, Selective Mode Conversion and Rainbow Trapping via Graded Elastic Waveguides, Phys. Rev. Applied 16, 034028 (2021).\\} 
The final publication is available at:
\textcolor{blue}{https://journals.aps.org/prapplied/abstract/10.1103/PhysRevApplied.16.034028}
\end{large}
% \preprint{APS/123-QED}
\clearpage

\title{Selective mode conversion and rainbow trapping via graded elastic waveguides}

\author{Jacopo M. De Ponti$^{1}$, Luca~Iorio$^{1}$, Emanuele Riva$^2$, Raffaele Ardito$^{1}$, Francesco Braghin$^2$, Alberto Corigliano$^1$}

\affiliation{$^1$ Dept. of Civil and Environmental Engineering, Politecnico di Milano, Piazza Leonardo da Vinci, 32, 20133 Milano, Italy \\
$^2$ Dept. of Mechanical Engineering, Politecnico di Milano, Via Giuseppe La Masa, 1, 20156 Milano, Italy \\
}

\begin{abstract}
We experimentally achieve wave mode conversion and rainbow trapping in an elastic waveguide loaded with an array of resonators. Rainbow trapping is a phenomenon that induces wave confinement as a result of a spatial variation of the wave velocity, here promoted by gently varying the length of consecutive resonators. 
By breaking the geometrical symmetry of the waveguide, we combine the wave speed reduction with a reflection mechanism that mode-converts flexural waves impinging on the array into torsional waves travelling along opposite directions. 
The framework presented herein may open opportunities in the context of wave manipulation through the realization of structural components with concurrent wave conversion and energy trapping capabilities.\\

\end{abstract}

\maketitle

\section{Introduction}
The emergence of mechanisms to manipulate the propagation of waves has attracted growing interest across different realms of physics, with multiple realizations in quantum \cite{hasan2010colloquium}, electromagnetic \cite{lu2014topological,khanikaev2013photonic}, acoustic \cite{cummer2016controlling,ma2016acoustic,yang2015topological} and elastic systems \cite{nassar2020nonreciprocity,wang2020tunable,hussein2014dynamics,Guenneau2009,Maurel2006,Deymier2003,Papanikolaou2011,Adams2009,Erturk2016,Kuperman2015,Erturk2018}.
In the context of mechanics, a number of works have recently investigated the emergence of non-trivial topological phases in elastic structures, in analogy with relevant behaviours previously observed in quantum physics \cite{huber2016topological}. Notable examples extensively explored in mechanics include defect-immune and scattering-free waveguides, which have been conceived in analogy to the quantum Hall (QH) \cite{wang2015topological,chen2019mechanical}, quantum spin Hall (QSH) \cite{susstrunk2015observation,chen2018elastic,chaunsali2018subwavelength,miniaci2018experimental}, and quantum valley Hall (QVH) effects \cite{riva2018tunable,vila2017observation,liu2018tunable,liu2019experimental}. Other approaches to topology-based design leverage 1D structures augmented by a virtual dimension in parameters space, to access topological properties typically attributed to 2D systems to pursue pumping of elastic waves \cite{rosa2019edge,riva2020edge,grinberg2020robust,riva2020adiabatic,cheng2020experimental,Ruzzene2020} and nonreciprocity \cite{marconi2020experimental,attarzadeh2020experimental}, to name a few. 
In other words, elastic waveguides are excellent candidates to explore physical phenomena, especially due to the abundance of supported modes with distinct polarizations and coupling among them, which can be relatively simple to be established as compared to the electromagnetic counterparts. While the existence of multiple modes that can hybridize in presence of asymmetry is generally undesired and difficult to grasp, in many cases the break of symmetry creates opportunities by lifting the accidental degeneracies. Mechanical systems with broken symmetries along the thickness direction can be employed for the nucleation of a double Dirac cone dispersion that features coupling between otherwise degenerate states, which is the key to emulate the spin orbital coupling in QSH-based waveguides \cite{miniaci2018experimental}. Undulated structures have been shown to exhibit frequency gaps and wave directionality due to coupling between axial and flexural vibrations \cite{Ruzzene2015,Ruzzene2016}. Efficient mode conversion between flexural and longitudinal waves has been achieved through trapped modes with perfect mode conversion (TMPC) in a quasi-bound state in the continuum (BIC) \cite{Cao2021}, or the transmodal Fabry-Pérot resonance \cite{Kweun2017}. Other approaches to induce modal coupling leverage nonlinear dynamics which, however, can be often unpractical due to the large amplitudes required to activate sufficiently strong nonlinear interactions \cite{Gonella2018,Gonella2015}. As reported by the aforementioned examples, the additional complexity induced by the symmetry break often reflects in relevant behaviours that are relatively unexplored in the field of mechanics. \\
A recent line of work employs a graded array of resonators embedded in a host structure to manipulate wave propagation by taking advantage of the resonator-structure interaction.
This modulation strategy promotes a wavenumber transformation that, in turn, activates a spatial decrease of the wave velocity at distinct frequencies, which is blueprint of the so called \textit{rainbow effect} \cite{DePonti2019, DePonti2020, DePonti2021, Erturk2021,Chaplain2020b}. Originally discovered in electromagnetic systems in non-uniform, linearly tapered, planar waveguides with cores of negative index material \cite{Hess2007}, the rainbow effect has been pursued within different research fields and through numerous physical platforms, among which acoustic systems \cite{Zhang2013, Staliunas2013, Garcia-Raffi2014}, water waves \cite{Craster2018}, and fluid loaded elastic plates \cite{Colquitt2018}. 
Similar configurations have been combined with deep elastic substrates to induce conversion between Rayleigh ($R$) into Shear ($S$) or Pressure ($P$) waves \cite{Colombi2016, Colombi2017, Colquitt2017, Chaplain2019}.
Despite the underlying physics is driven by the spatial variation of the wave speed, the existence of distinct wave modes and the nature of the coupling among them delineate a transition between reflection and trapping mechanisms \cite{Chaplain2020}. The former is induced by resonances and leads to wave scattering at the boundary of the first Brillouin zone. The latter is instead promoted by the coupling between crossing wave modes, typical of the second Brillouin zone, and hereafter employed to achieve concurrent wave mode-conversion and trapping.
While the majority of the studies are focused on a single phenomenon, the coexistence of multiple functionalities is not common, especially in the field of mechanics.\\
Motivated by this and by the opportunity to simultaneously achieve trapping and mode-conversion, in this paper we consider an elastic waveguide loaded by a graded array of resonators with variable length. This gradual variation promotes rainbow trapping of flexural waves throughout the array. The considered implementation embodies a broken symmetry with respect to the shear center of the waveguide which, in turn, activates a mode-conversion between impinging flexural waves and torsional waves traveling in the opposite direction through a phenomenon that is known as \textit{mode locking} in mechanics \cite{Manconi2012}. The numerical and experimental results presented herein demonstrate how a simple structure can be used to confine and mode convert elastic waves, expanding the range of possibilities in the context of wave manipulation and control, with implications of technological relevance for applications involving mechanical vibrations, such as nondestructive evaluation, ultrasonic imaging, and energy harvesting. For instance, flexural and torsional wave modes provide different resolution for nondestructive evaluation and testing, especially due to the associated wavelengths which allow for identification of cracks and defects of different shapes and dimensions \cite{Chimenti1997}. Also, different wave propagation properties (both in terms of wave speed and dispersion) yields different behavior in the context of short and long-range inspection \cite{Wilcox2001} of mechanical structures \cite{Ramatlo2020} and biomedical systems \cite{Clement2004,White2006,Hynynen2005}. A more straightforward application concern energy harvesting and enhanced sensing, as extensively discussed in a number of papers \cite{DePonti2019,DePonti2020,DePonti2021,Chaplain2020,Chaplain2020b,Erturk2021}. \\
The paper is organized as follows. In section II, a simplified model for the analysis of symmetry broken waveguides is presented and employed to tailor a suitable graded profile of locally unbalanced resonators. The theoretical aspects of the trapping and conversion mechanisms are unfolded by way of a dispersion analysis, which is preparatory for numerical end experimental analysis pursued in section III. Concluding remarks are presented in section IV.

\section{Simultaneous trapping and mode conversion: simplified modeling and design}

\begin{figure}[ht]
\centering
\includegraphics[width=0.48\textwidth]{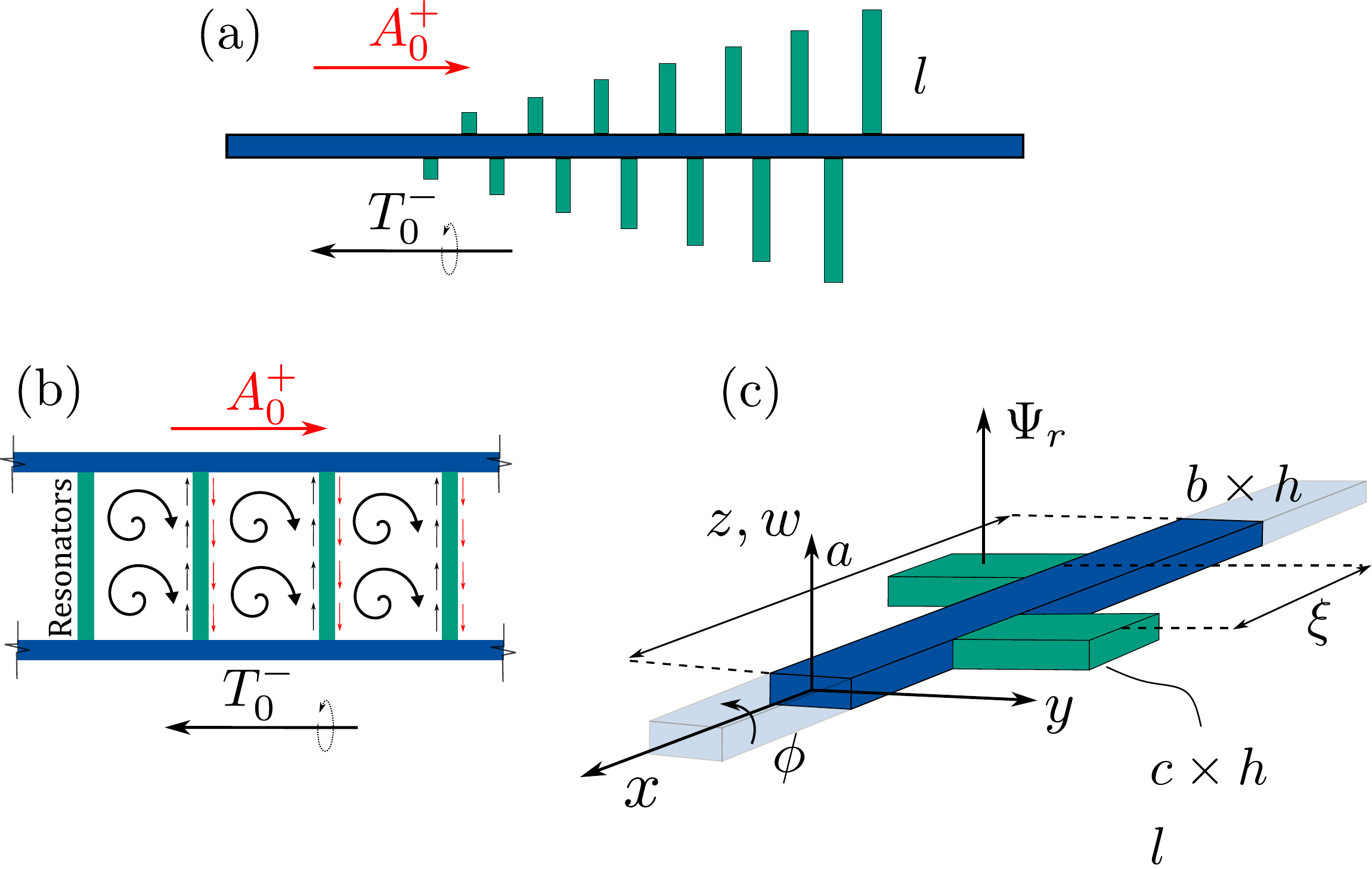}
\caption{(a) Graphical representation of the waveguide, which is made of a homogeneous beam rigidly connected to a graded array of resonators that activates the rainbow effect. Thanks to the broken geometrical symmetry, the waveguide operates as a converter between input $A_0^+$ and output $T_0^-$ wave modes. (b) Idealization of the conversion mechanism: the system can be conceptually interpreted as two distinct elastic waveguides supporting $A_0$ and $T_0$ wave modes, which are coupled using an array of resonators. (c) Schematic of the reference cell, that consists in identical resonator pairs separated by a non null distance $\xi$, along wih the employed reference system.  }
\label{fig:01}
\end{figure}

To elucidate the design paradigm and conditions for selective trapping and reflection, we consider the waveguide displayed in Figure \ref{fig:01}. The implementation consists in a slender beam, which is equipped with a set of attachments of variable length linearly increasing along the main dimension of the waveguide, to activate the rainbow effect for the asymmetric Lamb $A_0$ (input) wave mode. 
Each side of the beam accommodates an identical graded array of elements which, in turn, is rigidly connected to the host structure. In addition to the graded profile, the two arrays are spatially shifted by a distance $\xi$, to locally break the geometrical symmetry with respect to the shear centre of the cross-section. Such a tailored symmetry-break drives a spatially growing coupling between transverse motion $w$ and rotation $\phi$ of the waveguide, which are inherently linked to the asymmetric Lamb and torsional $T_0^-$ wave modes. This coupling activates a conversion mechanism that can be interpreted in the light of modal interaction between distinct waveguides, according to the schematic in Figure \ref{fig:01}(b). This model represents a conceptual scheme, while the theoretical derivations are obtained from the model in Fig. 1(a) and 1(c), respectively. The schematic illustrates how consecutive resonators locally trigger the transformation between $A_0$ and $T_0$ wave modes, inducing a reflected wave $T_0^-$ exiting from the array, as shown in Figure \ref{fig:01}(a) and further wave confinement due to multiple conversion between $T_0-A_0$ wave modes.

\begin{figure*}[t!]
\centering
\subfloat[]{\includegraphics[width=0.43\textwidth]{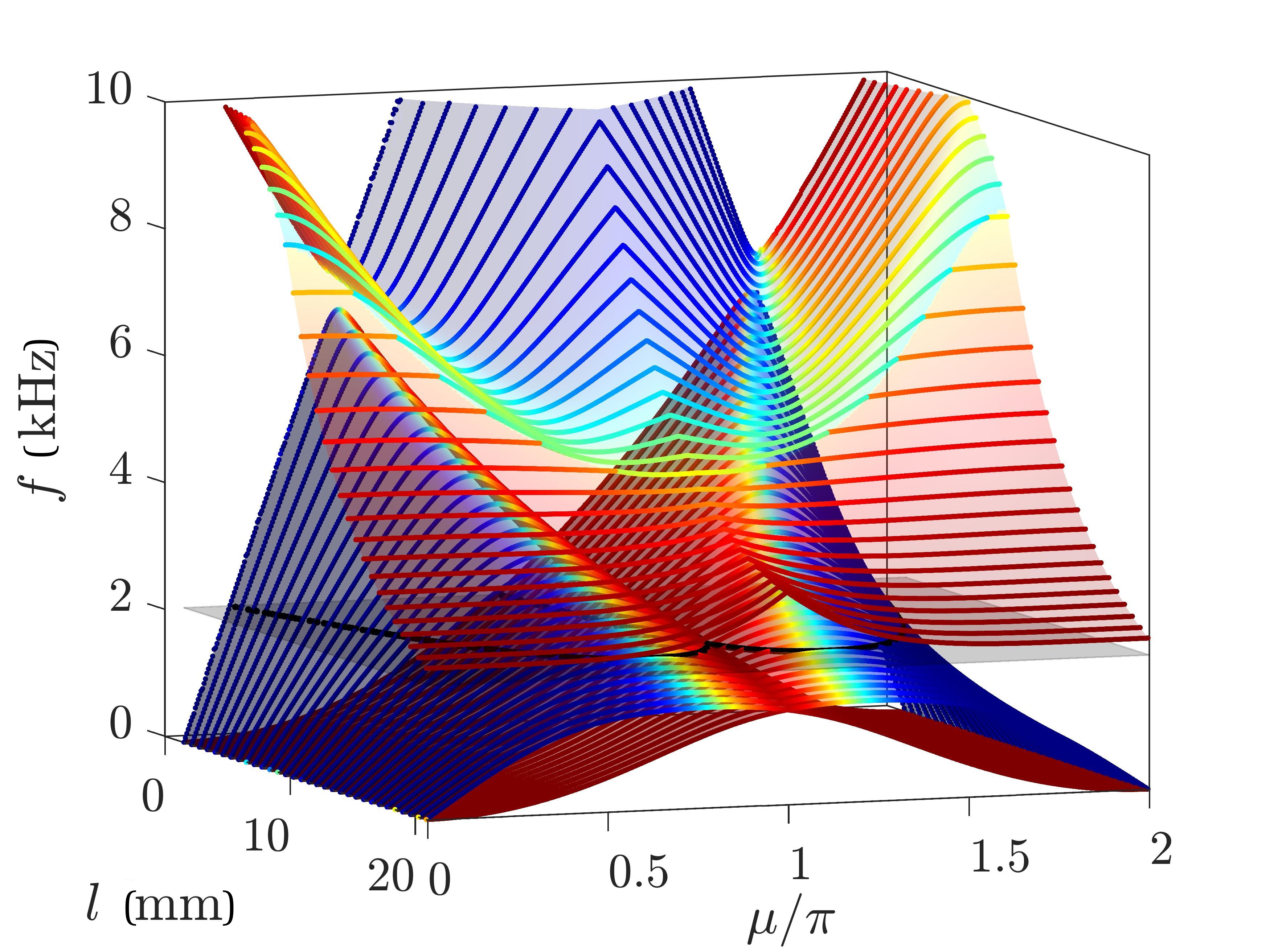}}
\subfloat[]{\includegraphics[width=0.43\textwidth]{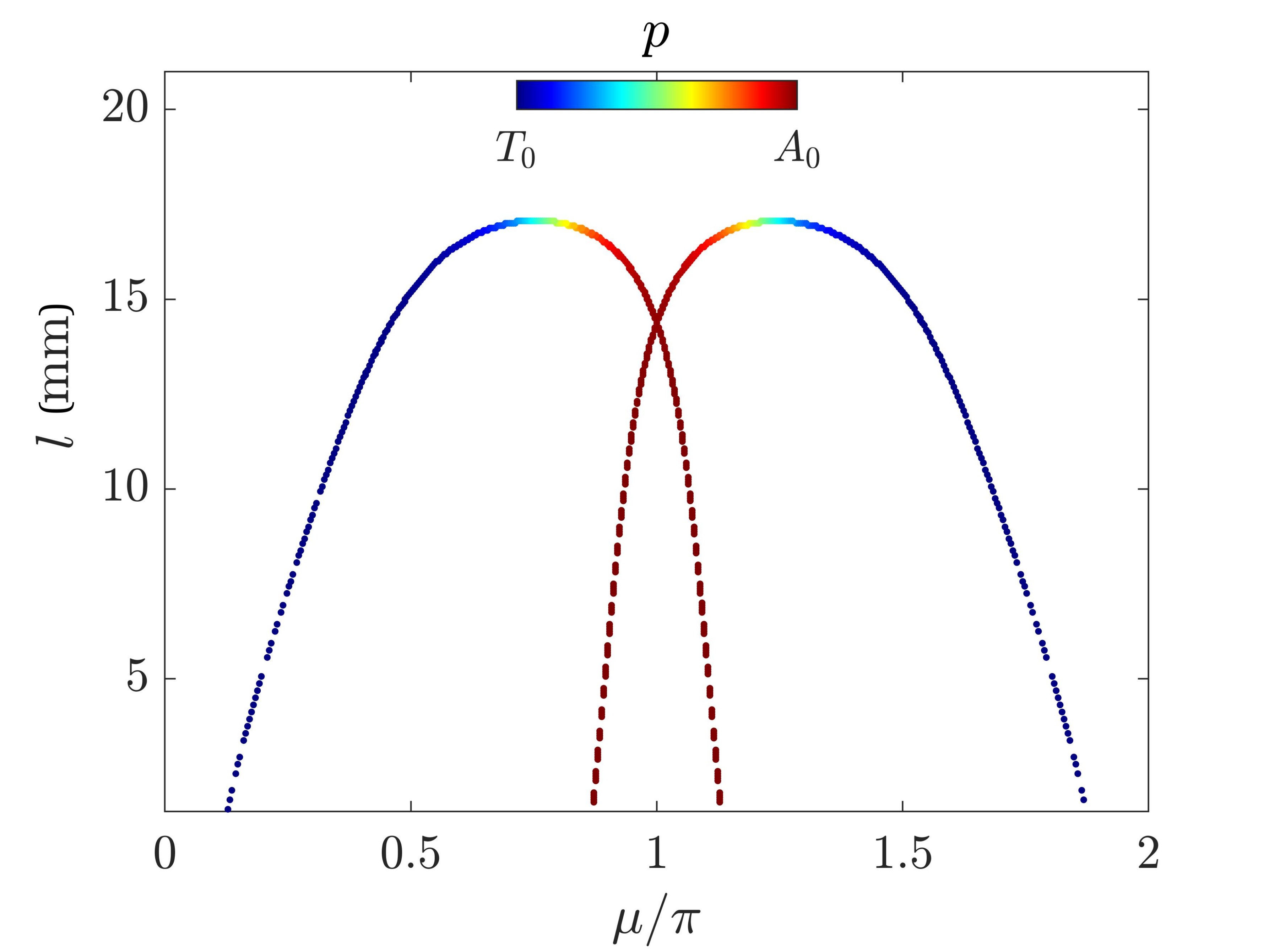}}\\
\subfloat[]{\includegraphics[width=0.375\textwidth]{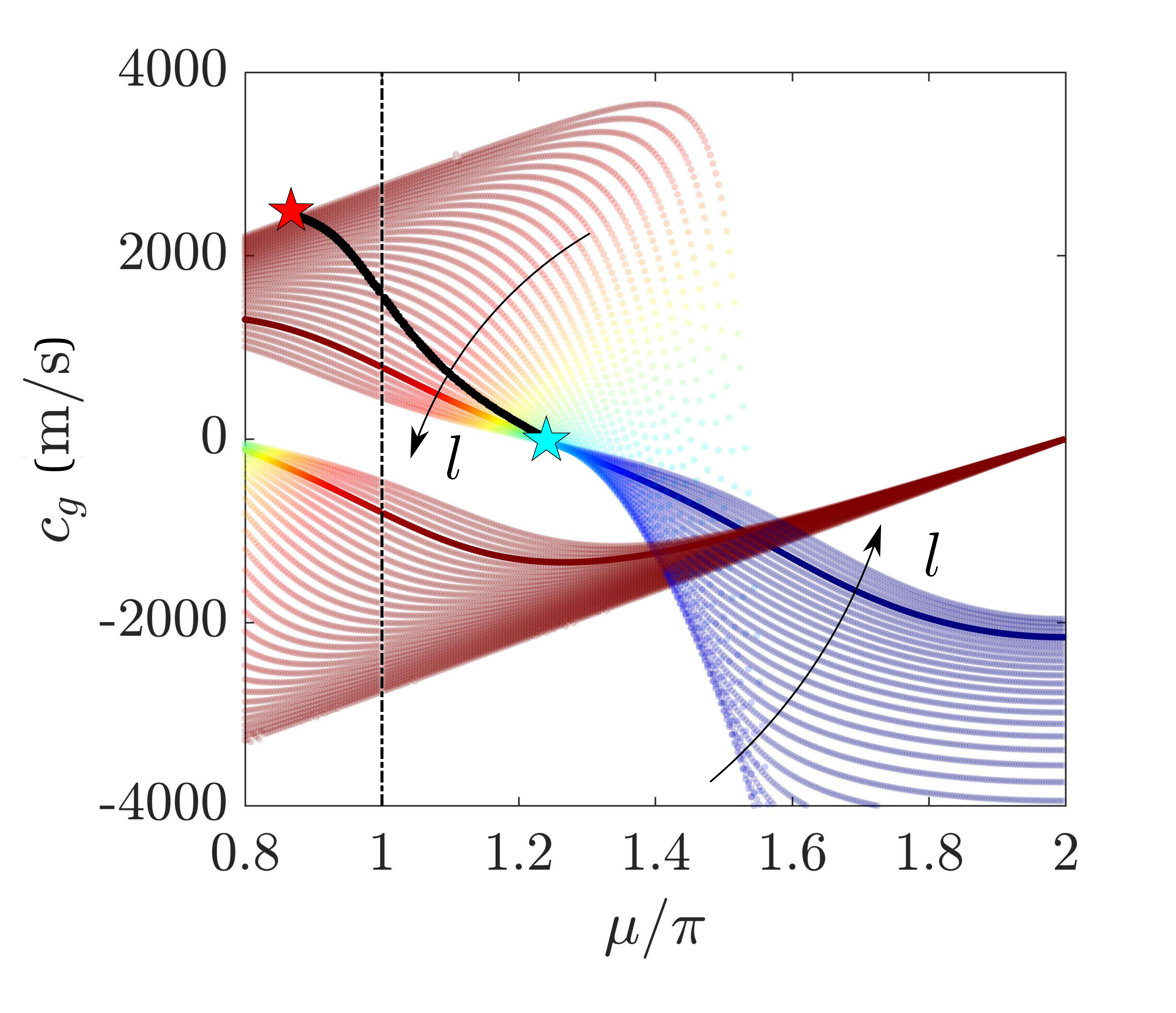}}
\subfloat[]{\includegraphics[width=0.62\textwidth]{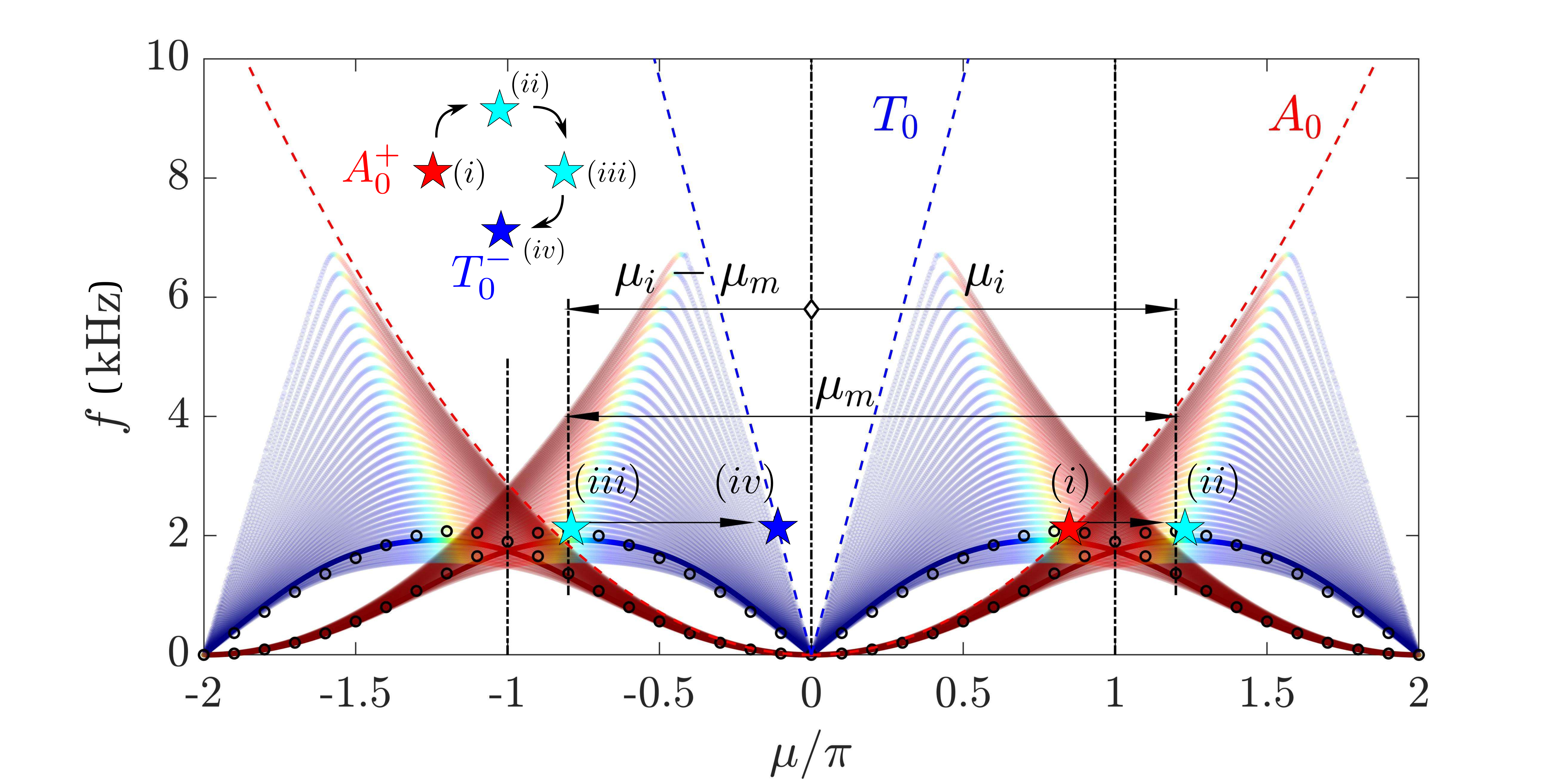}}
\caption{(a) Dispersion surface $\omega\left(\mu,l\right)/2\pi$. For ease of visualization, only the first three dispersion surfaces are displayed. The colour is proportional to the polarization factor $p$, which allows to discriminate between rotation $\phi$ and vertical displacement. Due to the broken symmetry and local resonance, the second and third dispersion surfaces are separated by a frequency gap. The isofrequency line, highlighted in black, represents the operating frequency of the structure, while the coloured version is reported in (b). (c) Group velocity $c_g$ upon varying the length of the resonators $l$ with change step $\Delta l = 0.315$ $mm$,  for the first two dispersion bands. In the figure, the group velocity line associated to the target dispersion branch is highlighted with darker dots. Such a target branch corresponds to the dispersion tangent to the isofrequency line displayed in (b). The group velocity $c_g\left(l\right)$ computed for the target frequency is superimposed to the coloured plot with black dots. (d) The dispersion relation $\omega\left(\mu\right)$ upon varying $l$. The schematic illustrate the conversion mechanism that undergoes in correspondence of the target resonator. The system is excited through the $A_0^+$ wave mode in correspondence to the red star (i), which is wavenumber transformed to the cyan star (ii), due to the grading. As the wave reaches the zero group velocity region within the second Brillouin Zone, it is backscattered to the cyan star (iii) by a quantity equal to the reciprocal lattice vector $\mu_m$, and wavenumber transformed to the back traveling $T_0^-$ wave mode marked with the blue star (iv). Black dots represent the numerical solution computed in COMSOL Multiphysics.}
\label{fig:02}
\end{figure*}

For simplicity, we assume that the further wave modes supported by the waveguide are orthogonal to the excitation mechanism and weakly coupled to the wave modes of relevance $A_0$ and $T_0$.
We also assume that the grading is gentle enough such that the local wave propagation properties within the array can be deduced from the dispersion curves of the constituent elements.
As such, the functional design and conceptual analysis of the grading profile are carried out considering the elastodynamic equations for an Euler-Bernoulli beam dominated by torsional-flexural coupling and loaded with periodic resonators \cite{achenbach2012wave}:
%\begin{equation}
%\resizebox{.99\hsize}{!}{$
%\begin{aligned}
\allowdisplaybreaks
\begin{small}
\begin{align}
    EI_y&w_{,xxxx}+\rho Aw_{,tt}=
    {\sum_{r=1}^{2}\sum_{n=-\infty}^{+\infty}}m\omega_0^2[\Psi_r\left(x_r+na\right)+\nonumber\\[4pt]
    &-w\left(x_r+na\right)+(-1)^r\frac{b+l}{2}\phi_r\left(x_r+na\right)]\delta\left(x-x_r-na\right)\tag{1a}\\[6pt]
    GJ_t&\phi_{,xx}-\rho J_p\phi_{,tt}=-{\sum_{r=1}^{2}\sum_{n=-\infty}^{+\infty}}(-1)^rm\omega_0^2\frac{b+l}{2}[\Psi_r\left(x_r+na\right)+\nonumber\\[4pt]
    &-w\left(x_r+na\right)+(-1)^r\frac{b+l}{2}\phi_r\left(x_r+na\right)]\delta\left(x-x_r-na\right)\tag{1b}\\[6pt]
    \Psi_{r_,tt}&\left(x_r+na\right)+\omega_0^2[\left(-1\right)^r\frac{b+l}{2}\phi\left(x_r+na\right)+\nonumber\\[4pt]
    &-w\left(x_r+na\right)+\Psi_r\left(x_r+na\right)]=0.\tag{1c}
%\end{aligned}%$}
\label{eq:01}
%\end{equation}
\end{align}
\end{small}

where $(\cdot)_{,x}$ denotes the partial derivative $\partial(\cdot)/\partial x$. $w$ and $\phi$ are the vertical displacement and the rotation of the beam, $\Psi_r$ is the displacement of the $n^{th}$ resonator in the absolute reference frame. $a=40\;{\rm mm}$ is the lattice size, $EI_y$, $\rho A$, $GJ_t$, and $\rho J_p$ are the elastic and inertial parameters associated to flexural and torsional motion of the beam, consistently with the unit cell and reference frame displayed in Figure \ref{fig:01}(c). $E=70\;{\rm GPa}$ and $\rho=2710\;{\rm kg/m^3}$ are the Young's modulus and the material density; $b=7\;{\rm mm}$ and $h=2\;{\rm mm}$ are the width and thickness of the host waveguide, respectively.
For simplicity, the dynamic contribution of the resonator pair is approximated in terms of equivalent bending stiffness $k$, participating mass $m$, and resonance frequency $\omega_0=\sqrt{k/m}$, which are dependent upon the resonator geometry $c,\;l$, and $h$.
The Dirac delta function $\delta\left(x-x_r-na\right)$ locally accounts for the presence of the resonators, placed at a distance $x_r=a/2\pm\xi/2$ from the left boundary of the lattice. Additional details on the simplified model are reported in the supplementary material \cite{SM}.\\
We investigate the dispersion properties $\omega\left(\mu,l\right)$ of the waveguide, where $l$ is considered as a free parameter and $\mu=\kappa_x a$ is the normalized wavenumber. To this end, we consider Ansatz for the displacement $w\left(x,t\right)=\hat{w}\left(x\right){\rm e}^{-j(\kappa_x x-\omega t)}$ and for the rotation $\phi\left(x,t\right)=\hat{\phi}\left(x\right){\rm e}^{-j(\kappa_x x-\omega t)}$ where $\hat{w}\left(x\right)=\sum_{p=-P}^{P}\hat{w}_p{\rm e}^{-jn\kappa_mx}$, $\hat{\phi}\left(x\right)=\sum_{p=-P}^{P}\hat{\phi}_p{\rm e}^{-jn\kappa_mx}$ embody the $x$-periodicity of the medium and $\kappa_m=2\pi/a$ is the modulation wavenumber. As such, the transverse and torsional motions are approximated in terms of $p=-P,\ldots,+P$ plane wave components. $P=3$ is found to be sufficient for an accurate description of the dynamic behavior at the operating frequency region.
Harmonic motion is also assumed for the resonators $\Psi_r\left(x_r\right)=\hat{\Psi}_r\left(x_r\right){\rm e}^{j\omega t}$. The application of the Plane Wave Expansion Method (PWEM), whose formulation is detailed in the supplementary material \cite{SM}, yields the following eigenvalue problem:

\begin{align}
    K\left(\kappa_x,l\right)\bm{\hat{\eta}}=\omega^2M\bm{\hat{\eta}
    }\tag{2}
    \label{eq:02}
\end{align}

where $K$ and $M$ are the $(2P+2)\times (2P+2)$ stiffness and mass matrices and $\bm{\hat{\eta}}=\left[\bm{\hat{w}},\bm{\hat{\phi}},\hat{\Psi}_1,\hat{\Psi}_2\right]^T$ accommodates the vector coefficients for the distinct wave modes and resonators pair. The solution of the eigenvalue problem $\omega\left(\kappa_x,l\right)$ is displayed in Figure \ref{fig:02}(a) for the first three dispersion bands. In the figure, the nature of the motion is discriminated through a color scale proportional to the polarization factor $p=|\int_0^a\hat{w}\left(x\right)|^2/\left(|\int_0^a\hat{w}\left(x\right)|^2+|b\int_0^a\hat{\phi}\left(x\right)|^2\right)$, which can be interpreted as a measure of the coupling between waves characterized by distinct polarizations. Thus, a transition from blue to red denotes a transformation from pure rotation to a transverse motion. 
Some considerations follow. (i) The lone beam features an accidental degeneracy, which corresponds to crossing flexural and rotational dispersion curves when $l\rightarrow0$. As the length of the resonator increases, the geometrical symmetry of the cross-section is locally broken, and a frequency gap emerges from the otherwise degenerate states; this phenomenon is generally known as \textit{mode locking} in mechanics, and is hereafter employed to tailor selective rainbow trapping and reflection. (ii) Interestingly, when the attachments length is smaller than a threshold (i.e. $l \approx 12\;{\rm mm}$), the bandgap formation mechanism is dominated by purely geometrical reasons; in contrast, for sufficiently high $l$ values, the resonance frequency of the attachment lies in the neighborhood of the gap, and the nature of the coupling is driven by a combination of broken symmetry and local resonance; practically, the combination of the two mechanisms yields a flattening of the dispersion curves that is beneficial in terms of wave velocity decrease and trapping. (iii) In the neighborhood of the gap, the dispersion relation exhibits wave modes characterized by a balanced torsional and flexural motion, especially in correspondence of the zero group velocity region ($a\partial\omega/\partial\mu=0$); this is of paramount importance for the interplay between wave conversion and energy trapping mechanisms, which will be discussed in the remainder of this section.  \\
The functional design of the graded profile of resonators is accomplished following the general guidelines provided in prior works \cite{DePonti2019, DePonti2021}. 
That is, rainbow trapping is hereafter pursued targeting a group velocity decrease along the beam at an operating frequency $f=2.12\;{\rm kHz}$. 
To that end, $l$ is denoted as the relevant parameter linearly varied along the main dimension of the beam. 
Such a variation yields a local wavenumber distribution $\mu\left(l\right)$ highlighted with black dots in Figure \ref{fig:02}(a)
The coloured version of the isofrequency line $\mu\left(l\right)$ is reported in Figure \ref{fig:02}(b) illustrating that a variation of $l$ not only activates wave modes characterized by different wavenumbers $\mu$, but also promotes a transformation between distinct polarizations (coloured dots). 
Consistently, the coloured group velocity profile $c_g\left(\omega(\mu),l\right)=a\partial\omega/\partial\mu$, is evaluated by finite difference for the entire wavenumber-parameter space and represented in Figure \ref{fig:02}(c) for the first two dispersion branches. For ease of visualization, the second dimension $l$ is eliminated and substituted with arrows, to better illustrate the group velocity profile $c_g\left(\omega(\mu)\right)$ in response to a variation of $l$. Here, the target dispersion curve, i.e. the dispersion branch that touches the isofrequency line with zero group velocity, is represented with darker-coloured dots.
As the length of the resonator $l$ is modified in space, the dispersion properties at the operating frequency naturally follow the black curve $\mu\left(l\right)$ displayed in Figure \ref{fig:02}(a). This modification is accompanied by a change in the group velocity $c_g\left(l\right)$ highlighted with black dots in Figure \ref{fig:02}(c) and, in turn, terminate in the zero group velocity region away from the edges of the Brillouin zone, which is the key factor to achieve rainbow trapping \cite{Chaplain2020}.
The analysis shows that the wave speed transformation is accompanied with a change in the polarization from a purely flexural mode that terminates into a region in which the wave is characterized by a mixed torsional-flexural motion and $c_g\approx 0$.

\begin{figure*}[t!]
\centering
\subfloat[]{\includegraphics[width=1.0\textwidth]{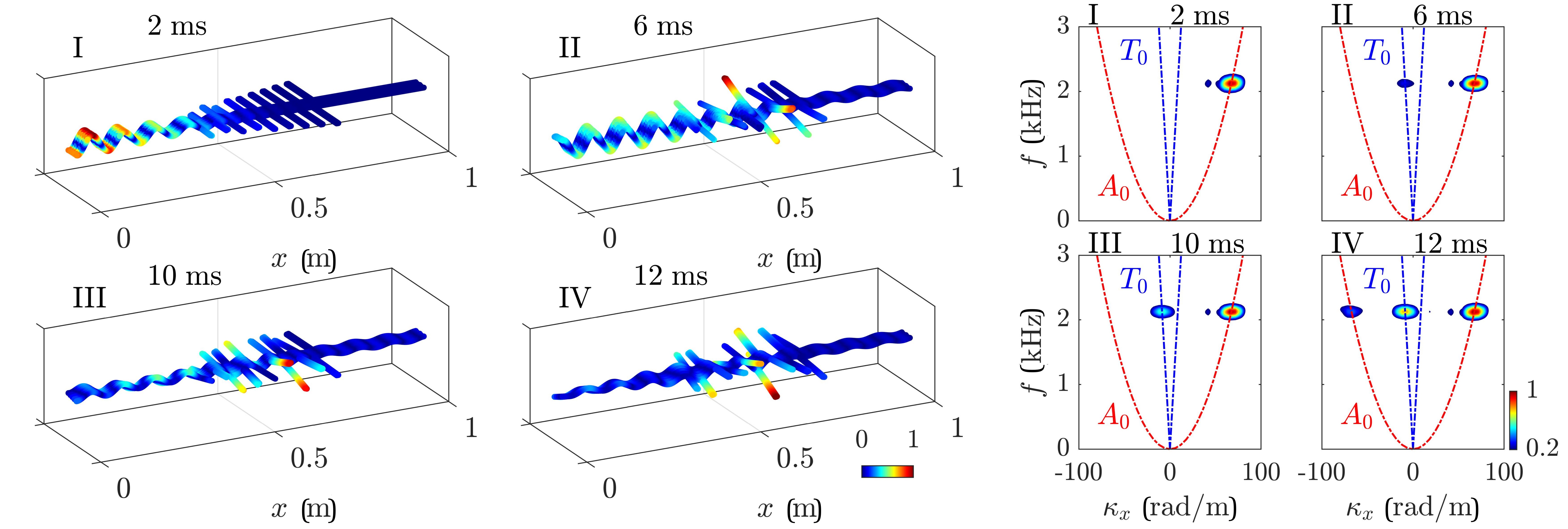}}\\
\subfloat[]{\includegraphics[width=0.32\textwidth]{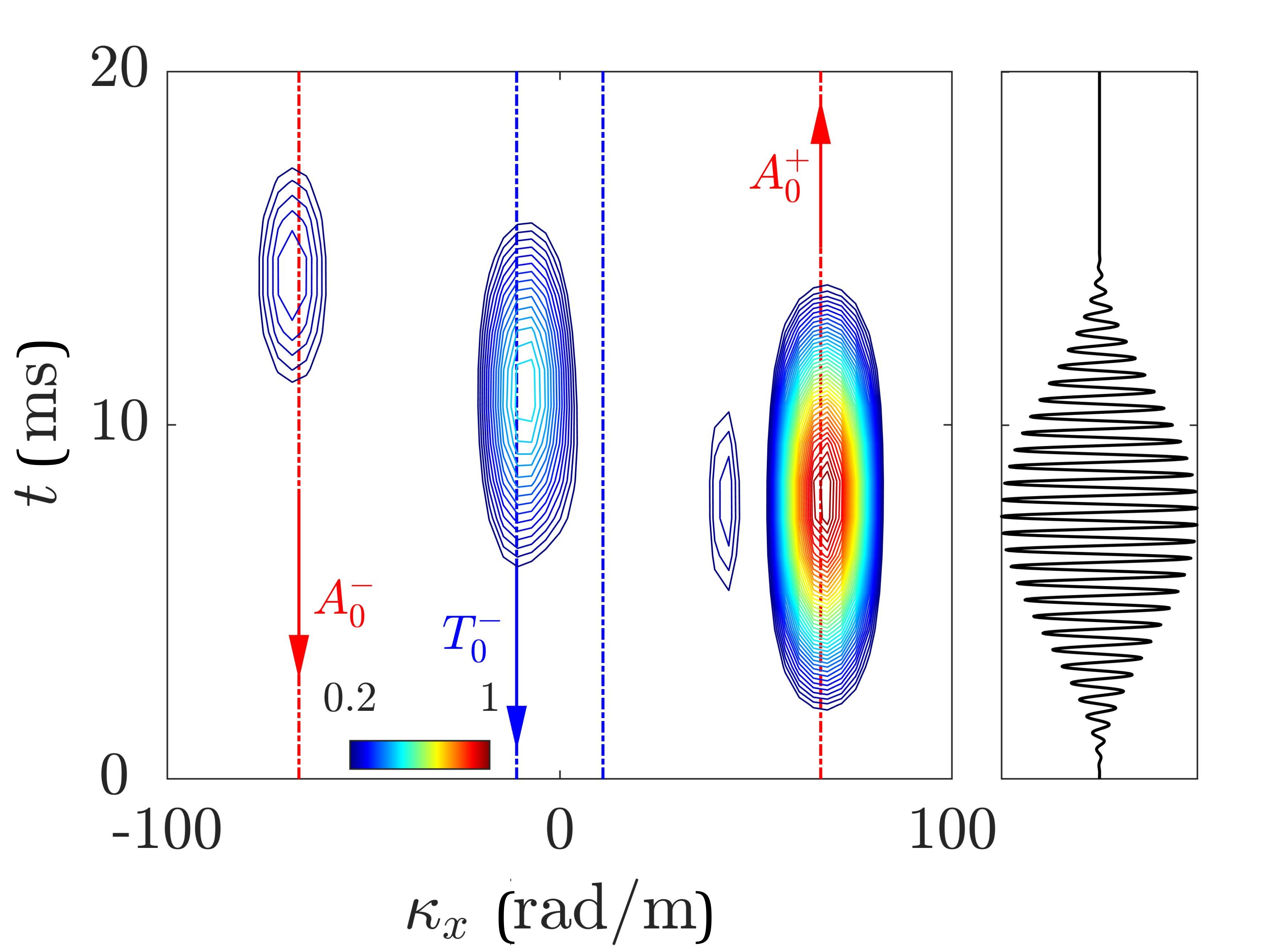}}
\subfloat[]{\includegraphics[width=0.32\textwidth]{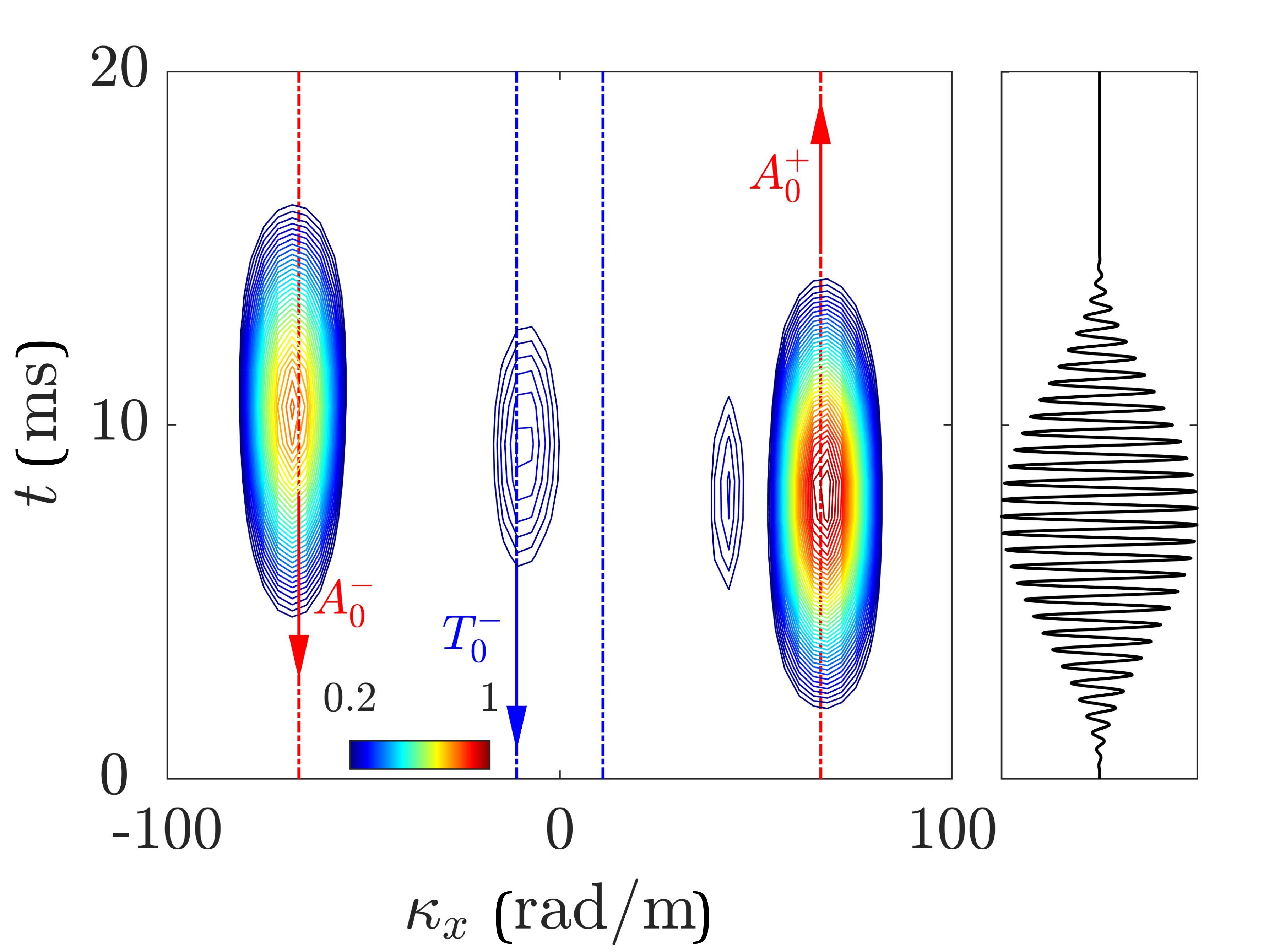}}
\subfloat[]{\includegraphics[width=0.32\textwidth]{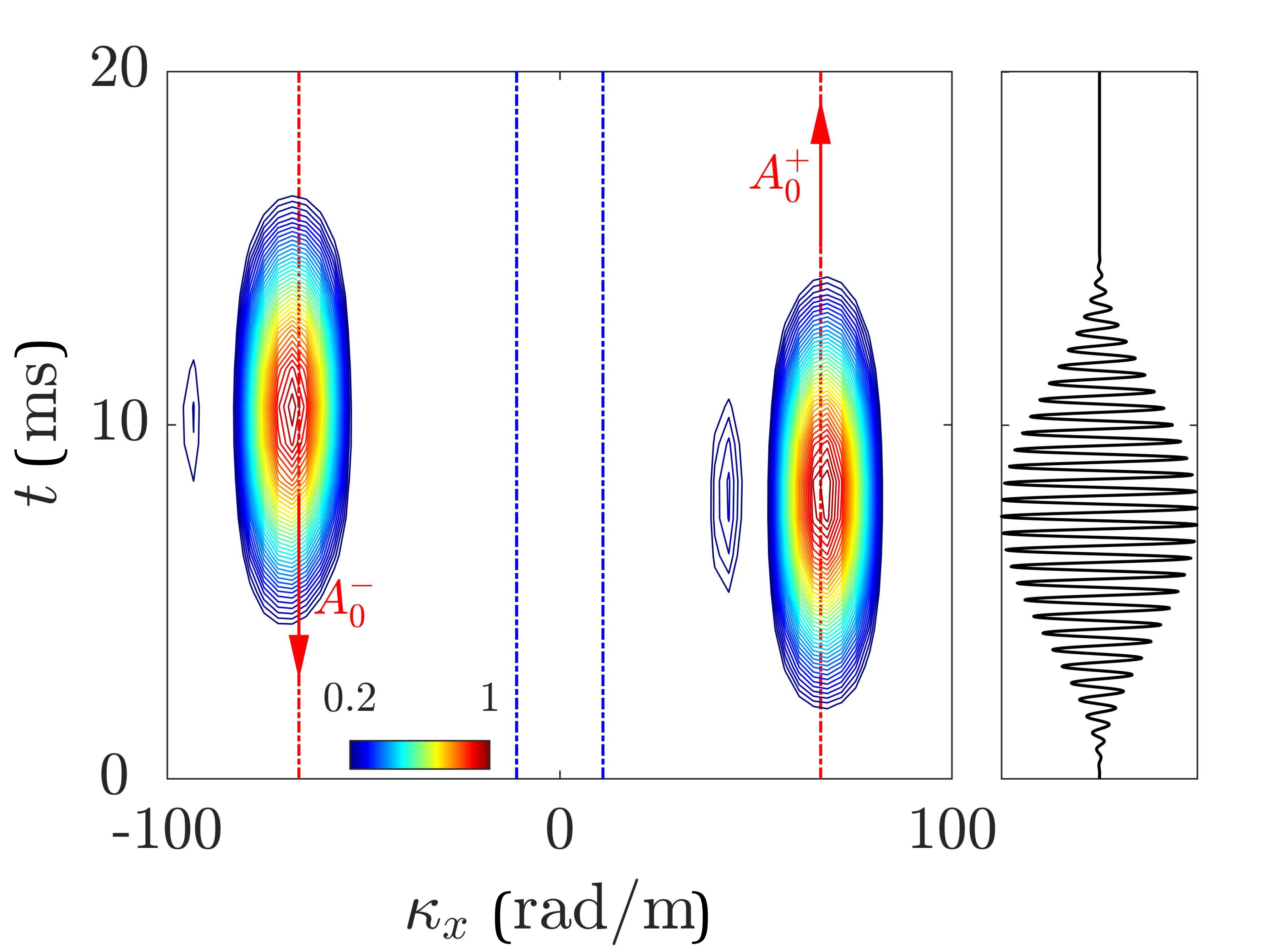}}\\
\subfloat[]{\includegraphics[width=0.32\textwidth]{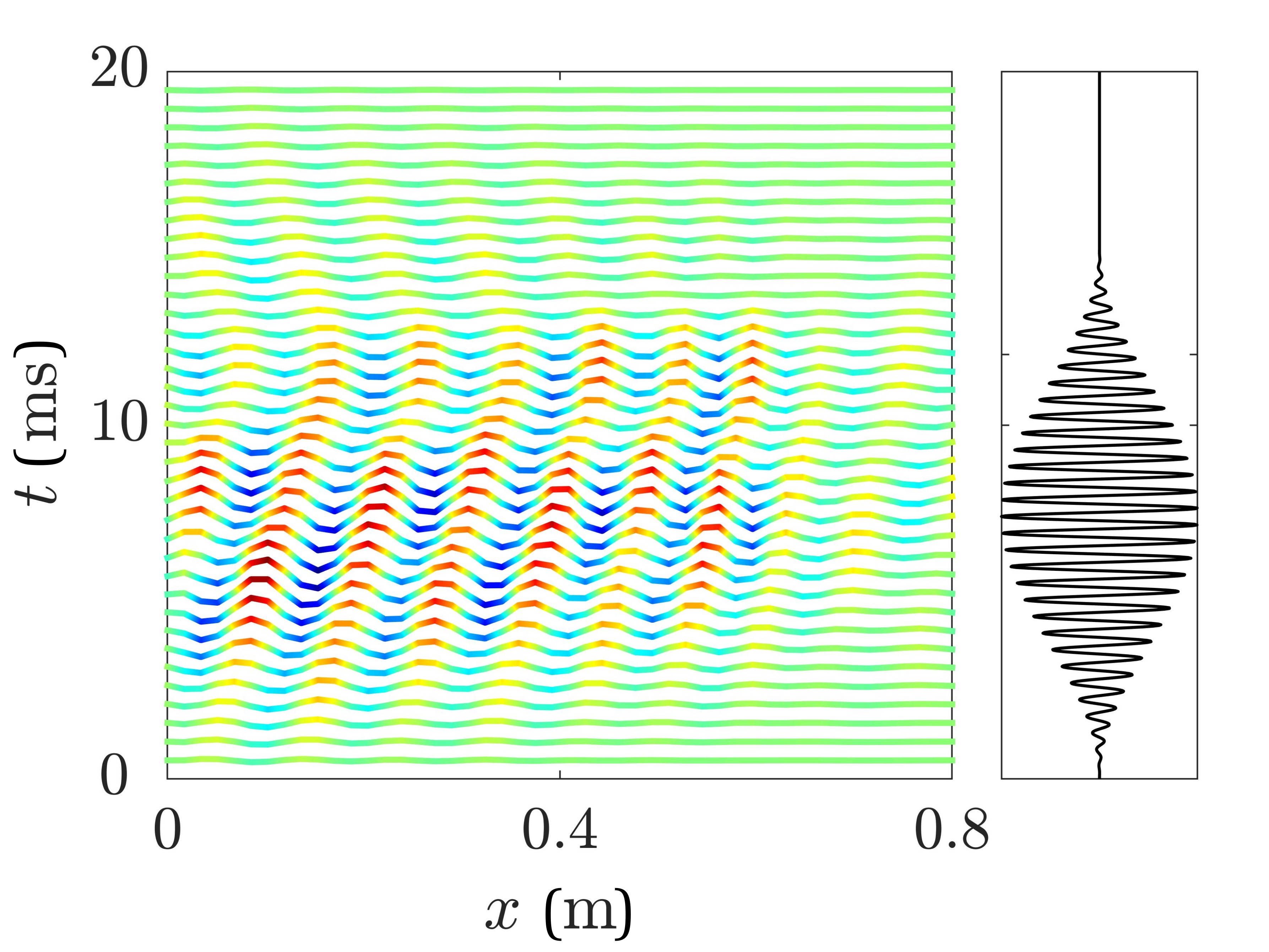}}
\subfloat[]{\includegraphics[width=0.32\textwidth]{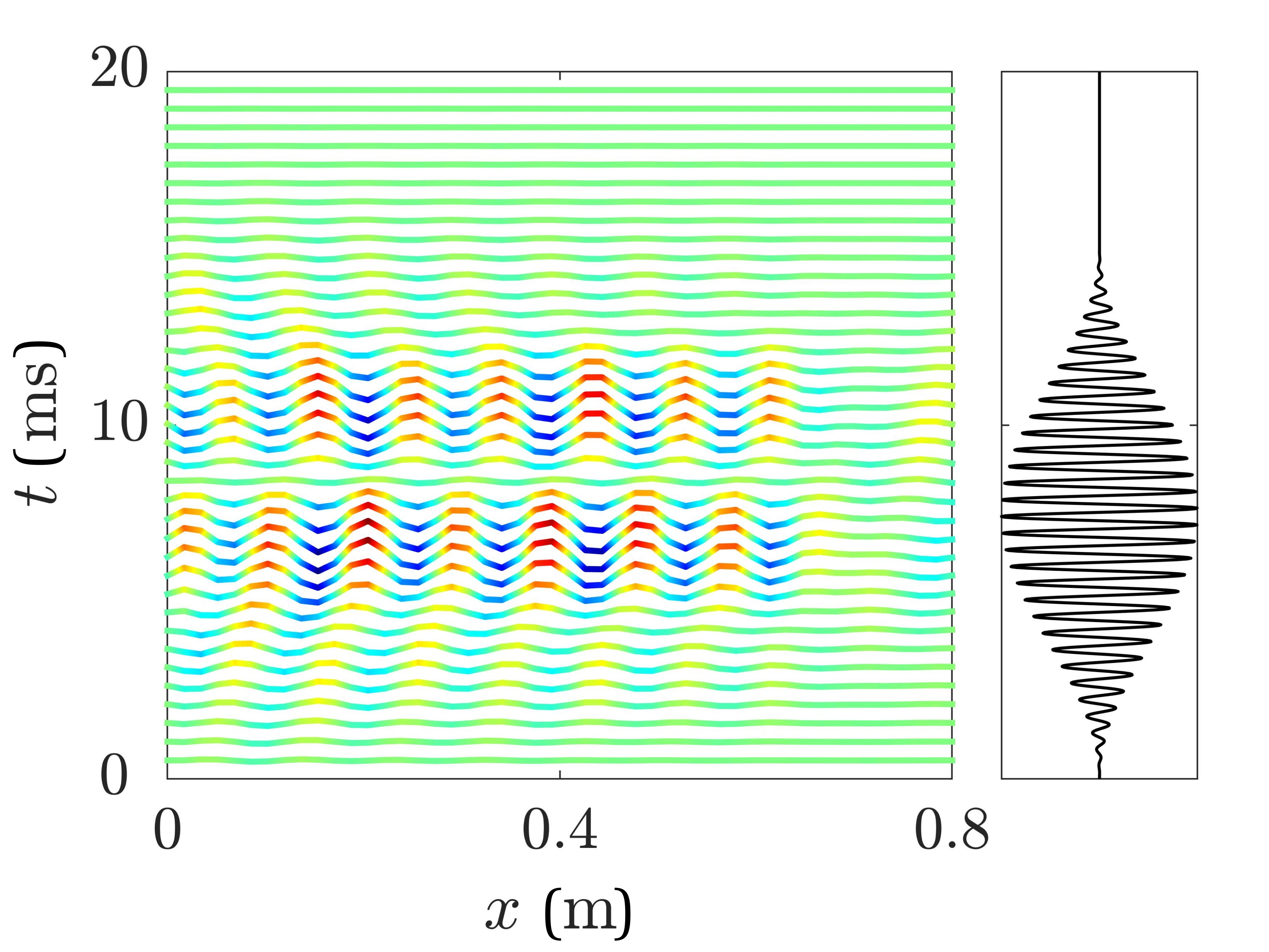}}
\subfloat[]{\includegraphics[width=0.32\textwidth]{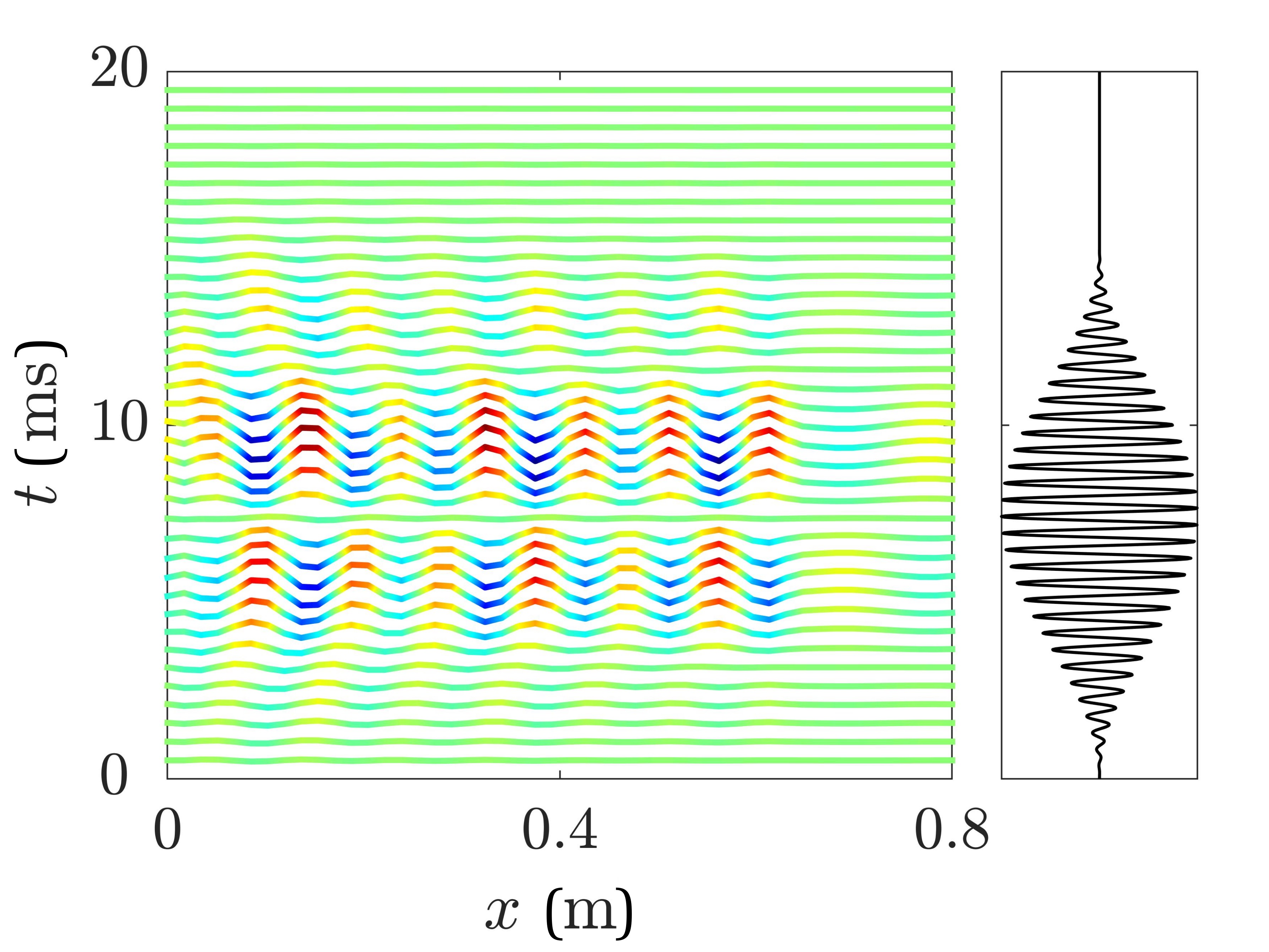}}
\caption{(a) Numerical displacement field measured at different time instants for an input wave mode $A_0^+$, along with the corresponding dispersion relations for a waveguide endowed with absorbing boundaries. The diagrams show that the input wave is reflected and converted from $A_0^+$ to $T_0^-$ in time. (b-d) The spectrograms display the temporal evolution of the spectral content for (b) a waveguide with the graded array of resonator; (c) a waveguide without the initial part of the grading, i.e. with only resonators $7-9$; (d) a homogeneous waveguide, without the array of resonators. The excitation signal is shown alongside the spectrogram. The input wave is confined, trapped and delayed only if the complete array of resonators is present. (e-g) Corresponding out-of-plane displacement field filtered to separate the contribution of the $A_0$ wave mode. }
\label{fig:03}
\end{figure*} 

In addition to the rainbow effect, the tailored broken symmetry of the array is responsible for a reflection mechanism that mode-converts the impinging flexural waves into torsional waves traveling along the opposite direction. This mechanism is hereafter discussed assuming that most of the conversion takes place in correspondence of the zero group velocity region, where the coupling is maximum. As such, among the possible dispersion branches, we focus on the (target) curve highlighted with darker-coloured dots in Figures \ref{fig:02}(c-d) and we assume that a smooth variation of $l$ is provided to reach the target curve.
Now, consider an input wave $A_0^+$, represented in figure \ref{fig:02}(d) with the red star $(i)$. Such a wave impinges on the graded array of resonators and, due to the smooth variation of $l$ in space, the wave mode experiences a wavenumber transformation that drives the transfer of spectral content to the cyan star $(ii)$, where the group velocity at the operating frequency is ideally nullified and the normalized wavenumber is $\mu=\mu_i$. It is worth to mention that the group velocity can reach a zero value only for infinitely long arrays, while for a finite number of elements the wave propagation occurs with decreased speed and, therefore, the positive-going component impinges on the following resonators. Due to the conservation of the crystal momentum, the traveling wave is back-scattered to $\mu=\mu_i-\mu_m$ and, once again, is wavenumber-transformed from the cyan star $(iii)$ to the blue star $(iv)$ as a result of the linear decrease of $l$, which produces a back-propagating torsional wave mode $T_0^-$ exiting the graded array. Due to the limited number of resonators, a small amount of energy doesn't follow this path, leading to a back-scattered flexural wave $A_0^-$ with spectral content $\mu=-\mu_i$, which is delayed in time consistently with the wave speed reduction described by the dispersion analysis. According with the above arguments and provided that only a small amount of energy doesn't follow such a wavenumber transformation in the back-propagating path, the conversion efficiency of such a system is close to $100\%$. A more detailed discussion on this matter is provided in the supplementary material 
\cite{SM}.\\
To conclude the design of the graded array, the numerical dispersion relation is evaluated numerically via COMSOL multiphysics environment spanning the parameter space $l\in[0,22]\;{\rm mm}$ and reported with black circles in Figure \ref{fig:02}(d) for the target dispersion curve. According to the analysis, the necessary array of resonators to reach near-zero group velocity is characterized by initial and final (target) length of $l_{1}=8\;{\rm mm}$ and $l_{7}=18.3\;{\rm mm}$, respectively, which are distributed along $L=7a\;{\rm mm}$ for a number of $N=7$ cells. The array is then continued to a final length of $l_{9}=21.2\;{\rm mm}$ to prevent wave propagation through the array for frequencies in the neighborhood of the target mode.\\

\section{Theory meets experiments}
The discussion is now focused on the transient analysis of wave propagation simulated in Abaqus implicit environment \cite{Implicit}, to corroborate the theoretical claims. The implementation consists in a homogeneous beam that serves as input domain for right traveling waves, which is followed by the graded array of resonators. Undesired reflections are avoided by way of absorbing boundaries applied to the left and right ends of the beam \cite{RAJAGOPAL201230}. A transverse force with central frequency $f_0=2.12\;{\rm kHz}$, width $\Delta f=0.14\;{\rm kHz}$ and time duration of $15\;{\rm ms}$ is employed to provide excitation and confine the energy content in the neighborhood of the operating frequency. 
Additional details on the numerical methods are reported in the supplementary material \cite{SM}.\\
Numerical results for consecutive time instants $t_0=2,6,10,12\;{\rm ms}$ are displayed in Figure \ref{fig:03}(a). As expected, at the beginning of the time simulation, the motion is dominated by flexural waves. As time elapses, part of the energy is transferred to torsional wave modes, starting from the snapshot captured in Figure \ref{fig:03}(a)II at $t_0=6\;{\rm ms}$, where the homogeneous trait is dominated by transverse waves and the graded array exhibits rotation, due to the broken geometrical symmetry. 
In contrast, the displacement field displayed in Figures \ref{fig:03}(a)III-IV is characterized by mixed wave motion within the entire spatial domain. These considerations are further confirmed by the numerical dispersion $\hat{d}\left(\kappa_x,\kappa_y,\omega\right)=\sqrt{\hat{w}^2+\hat{u}^2+\hat{v}^2}$, which is illustrated alongside the displacement fields in Figure \ref{fig:03}(a), where $\hat{w}=\hat{w}\left(\kappa_x,\kappa_y,\omega\right)$, $\hat{v}=\hat{w}\left(\kappa_x,\kappa_y,\omega\right)$, and $\hat{u}=\hat{w}\left(\kappa_x,\kappa_y,\omega\right)$ are the 3D Fourier-transformed displacement fields of the homogeneous part of the waveguide. To evaluate the numerical dispersion, the time histories $w\left(x,y,t\right)$, $v\left(x,y,t\right)$, and $u\left(x,y,t\right)$ are windowed in the neighborhood of the probed time instants $t_0$ by way of a suitable Gaussian function $g(t)={\rm e}^{-(t-t_0)^2/2c^2}$ where $c$ is a parameter that controls the width of the Gaussian function. For ease of visualization, the numerical dispersion is reduced to $\hat{d}\left(\kappa_x,\omega\right)$ by taking the Root Mean Square (RMS) value along $\kappa_y$ which yields the diagram displayed in the figure. As expected, the spectral content in the waveguide is initially located in the positive half of the reciprocal space and relates to the $A_0^+$ dispersion branch. When the flexural wave reaches the array, the wave is trapped and converted into a component traveling along the opposite direction, corresponding to the $T_0^{-}$ and the $A_0^-$ dispersion branches exiting the graded array. The concept is elucidated in the spectrogram displayed in Figure \ref{fig:03}(b), which is obtained by smoothly varying the position of the Gaussian function spanning the range $t_0\in\left[0,20\right] {\rm ms}$. Also the dependence on frequency is eliminated by taking the RMS value, which results in the amplitude $\hat{d}\left(\kappa_x,t_0\right)$. The same analysis is performed on a beam without the initial part of the array (see Figure \ref{fig:03}(c)), responsible for the wave speed reduction, and on a homogeneous beam without graded array of resonators (reported in Figure \ref{fig:03}(d)) equipped with clamps in correspondence of resonators $N=8,9$.

\begin{figure}[ht!]
\centering
\includegraphics[width=0.5\textwidth]{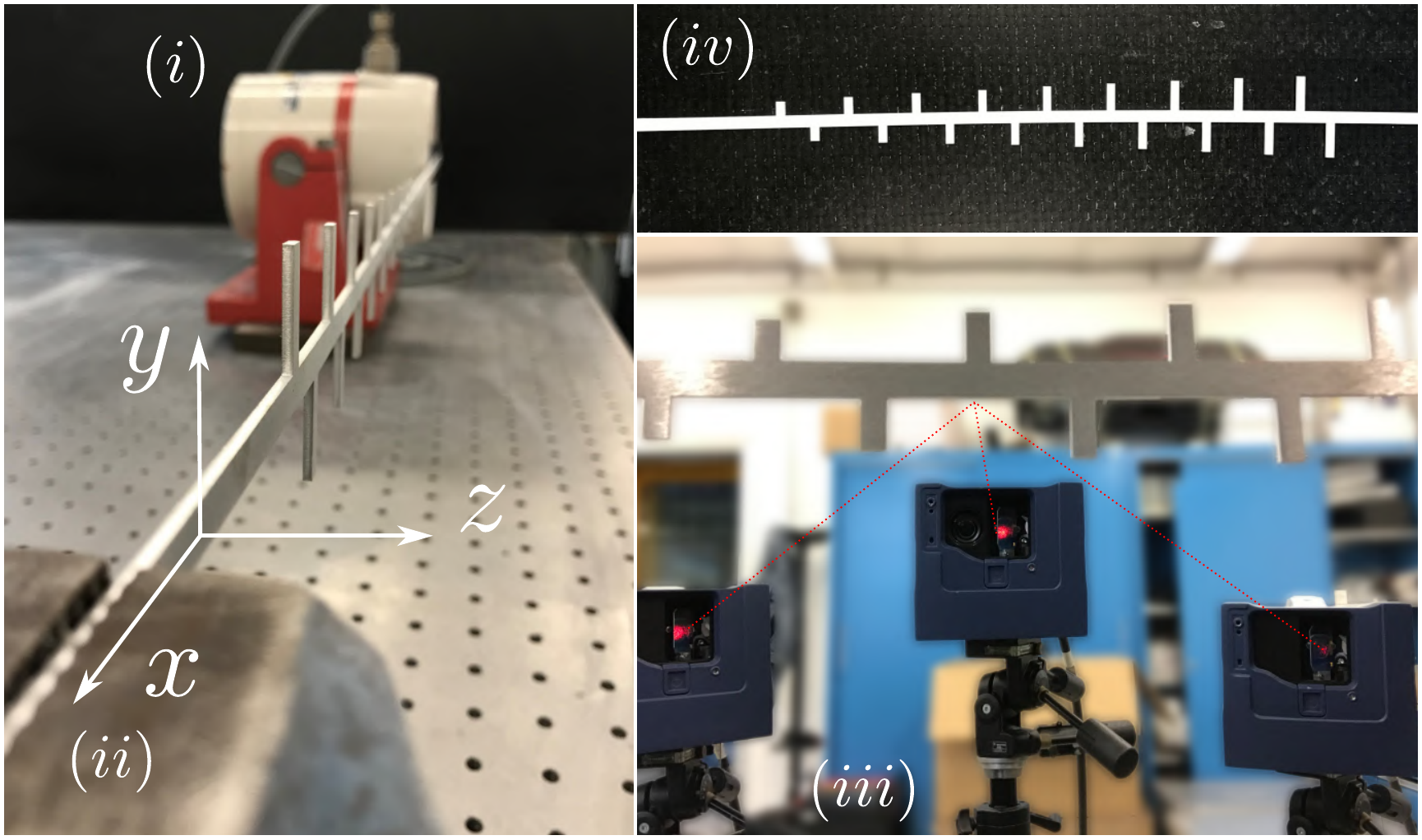}
\caption{Experimental setup employed to measure wave propagation. The waveguide is excited using an electrodynamic shaker ($i$), while the opposite side is clamped ($ii$). Wave propagation is measured on the surface of the waveguide through a 3D Polytec Scanner Laser Doppler Vibrometer ($iii$), which is able to separate the 3D velocity field. The top view of the graded array is also provided ($iv$). }
\label{fig:04}
\end{figure}

\begin{figure*}[t!]
\centering
\subfloat[]{\includegraphics[width=1.0\textwidth]{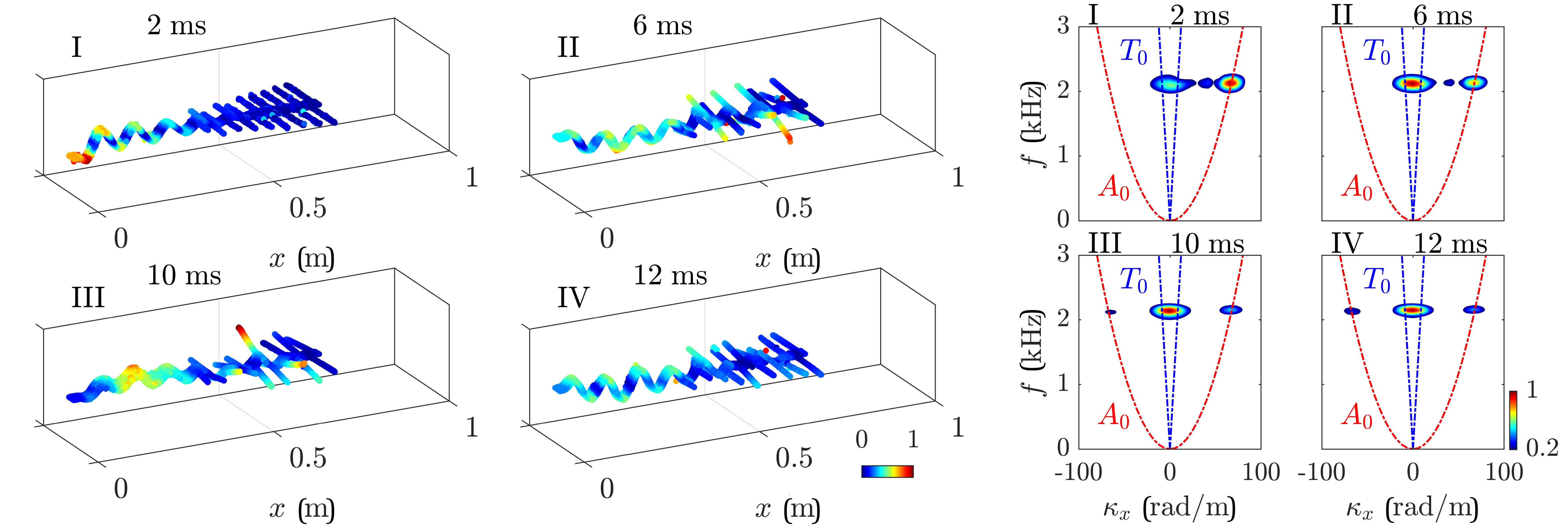}}\\
\subfloat[]{\includegraphics[width=0.32\textwidth]{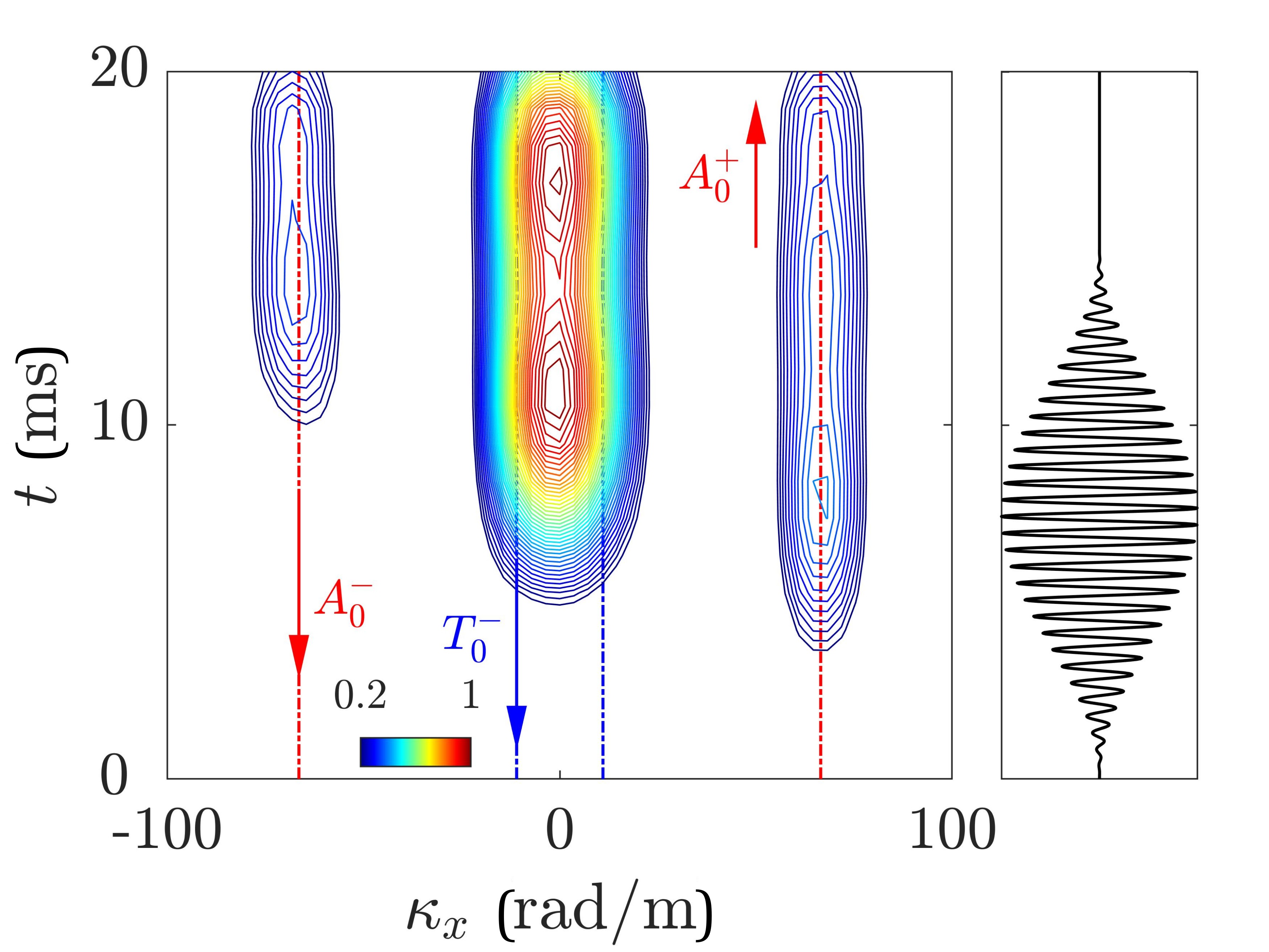}}
\subfloat[]{\includegraphics[width=0.32\textwidth]{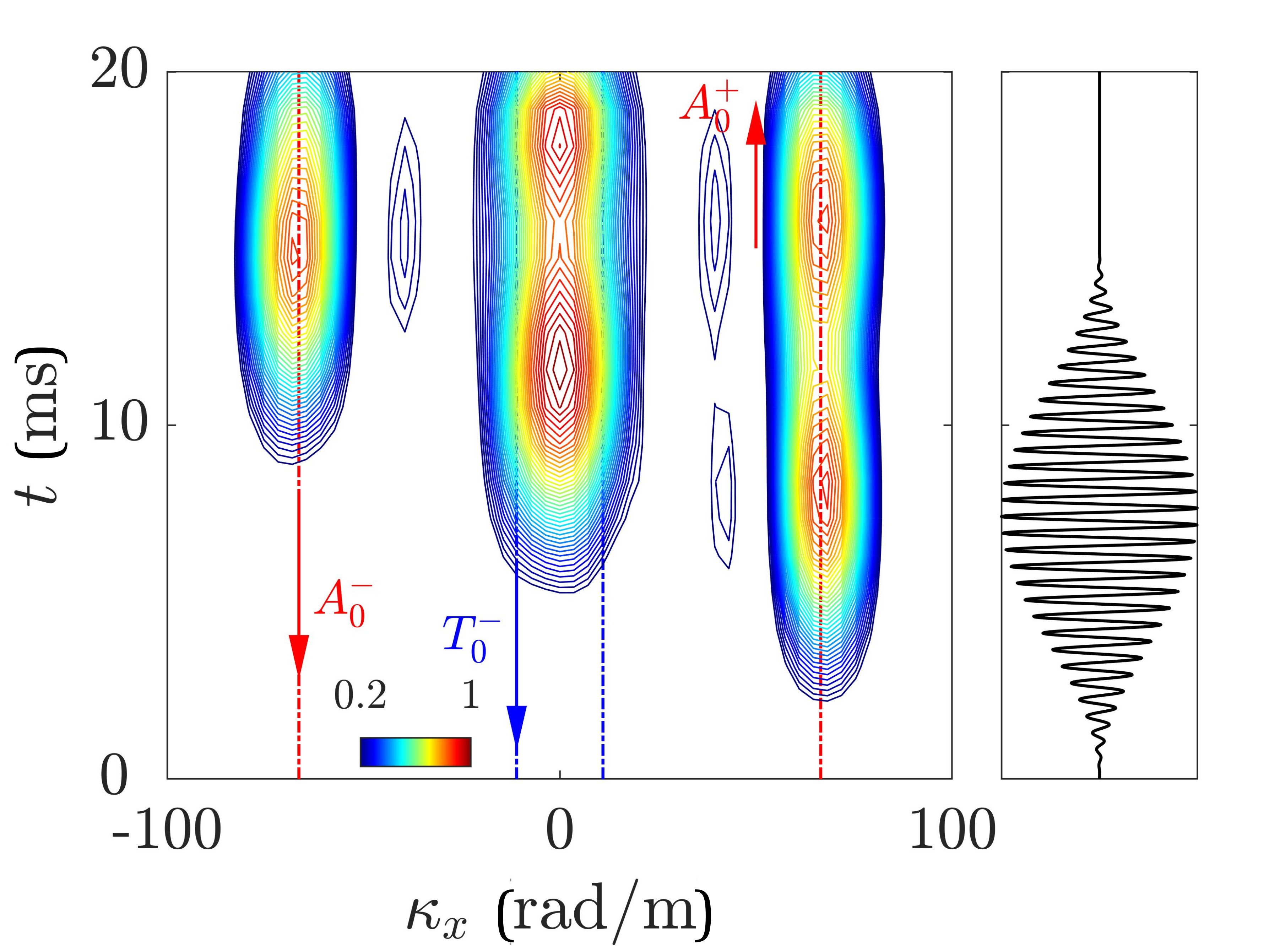}}
\caption{(a) Experimental velocity field and corresponding dispersion relation measured at different time instants. (b) Experimental and (c) numerical spectrogram, illustrating the evolution of the spectral content in time. Experimental and numerical results are characterized by a comparable temporal delay for the $A_0^-$ wave mode, while different conversion amplitudes are observed, due to the presence of imperfections and misalignments that enhance the torsional coupling.}
\label{fig:05}
\end{figure*}

A qualitative comparison of Figures \ref{fig:03}(b-d) reveals that: (i) the presence of the graded array shifts the backward propagating wave $A_0^-$ by $5$ ms, as a result of the tailored trapping mechanism, while the backward propagating $A_0^-$ component is not delayed in a relevant way in case the initial part of the array is not considered; (ii) the wave speed reduction is accompanied by a transfer of energy between $A_0^+$ and $T_0^-$ wave modes, which occurs approximately $3$ ms before the $A_0^-$ wave mode back-propagates; (iii) the amount of conversion is dramatically reduced in case the first part of the array is not present.
To further confirm that the graded array operates as delay line and mode converter for flexural/torsional waves, the displacement field $w\left(x,y,t\right)$ is filtered in wavenumber domain for $\kappa_x$ values outside $40\;{\rm rad/m}$ and $100\;{\rm rad/m}$. The filtered displacement field allows to isolate and graphically visualize the $A_0$ mode evolution in time, that is displayed in Figure \ref{fig:03}(e-g) for the three reference configurations. As expected, only if the graded array is present, the numerical wavefield is altered in a way that the energy slowly vanishes and is mode-converted to the counter-propagating $T_0^-$ wave mode. It is also worth noticing that the conversion efficiency is very high, as most of the energy, initially injected into $A_0^+$, is converted into the $T_0^-$ back-propagating mode.
The non-converted energy is instead back-reflected as a $A_0^-$ mode and shifted in time, consistently with the above discussion. Additional analysis on the conversion efficiency have been performed on structures with longer input domains and upon varying the number of pair resonators ($9$, $25$, and $50$). Such analysis are reported and discussed in the supplementary material \cite{SM} and demonstrate that for a sufficiently long system compliant with the theoretical arguments, the conversion efficiency is close to $100\%$.\\

The numerical results are validated through an experimental analysis in the transient regime, which is performed on a structure identical to the one employed in the simulations, except for the boundary conditions that are hereafter discussed. 
The schematic of the experimental setup is shown in Figure \ref{fig:04}. The system is rigidly connected to a LDS v406 electrodynamic shaker, to provide excitation, while the opposite end is clamped to ground, to avoid excessive geometrical deformation. The wavefield is measured on the surface of the system through a Polytec 3D Scanner Laser Doppler Vibrometer (SLDV), which is able to separate the 3D velocity field in both space and time. The excitation signal is synchronously started with the acquisition which, in turn, is averaged in time to decrease the noise. Additional information on the experimental methods are reported in the supplementary material \cite{SM}. The experimental results in terms of the 3D displacement field $w\left(x,y,t\right)$, $v\left(x,y,t\right)$, and $u\left(x,y,t\right)$, along with the corresponding dispersion $\hat{d}\left(\kappa_x,\omega\right)$ centered at different time instants $t_0$ are displayed in Figure \ref{fig:05}(a), which are similar to the ones displayed in Figure \ref{fig:03}(a). After the input $A_0^+$ mode is injected in the structure, part of the energy is mode-converted in a torsional $T_0^\pm$ mode and part is delayed in time. In contrast to the results displayed in Figure \ref{fig:03}(a), the propagating waves cannot escape the system, and the amount of dissipation is negligible. We also observe that, in comparison to the numerical case-study, the mode-conversion takes place in shorter time, and the associated amplitude is greater. This is attributed to: (i) the absence of absorbing boundaries, which facilitates the accumulation of energy, especially that transferred to the torsional wave mode characterized by faster propagation speed and triggers the formation of a standing mode; (ii) the presence of misalignment and imperfections that facilitate the conversion between wave modes that populate the dispersion at the operating frequency.\\
The same considerations apply for the experimental and numerical spectrograms in Figure \ref{fig:05}(b,c). The figures illustrate a similar amount of delay between incident $A_0^+$ and reflected $A_0^-,\;T_0^-$ wave modes, whereas in the experimental diagram in Figure \ref{fig:05}(b) we observe an extra conversion amplitude between $A_0$ and $T_0$ waves that is triggered by the undesired curvature of the beam, due to the manufacturing process that is not present in the numerical analysis. Also, consistently with the simulation performed with absorbing boundaries, (\ref{fig:03}(b)) the converted $A_0^-$ wave mode is delayed by the same amount as for the experimental data displayed in \ref{fig:05}(b), which further confirms that the rainbow trapping occurs as expected. We finally notice that an additional $T_0^+$ component is measured; this is attributed to the finite length of the system, which allows for spurious edge reflections of the $T_0^-$ waves exiting the array.

\section*{Conclusions}
In this manuscript we have experimentally realized a graded structure that is capable of converting flexural waves into torsional waves traveling along the opposite direction. Such a conversion is accompanied with trapping of flexural waves that are delayed while propagating through the array. The complex interplay between trapping and conversion is explained in terms of dispersion analysis, where the dynamic behavior is dominated by a local symmetry break and the smooth variation of the resonator length in space, which locally activates the energy transfer between distinct wave modes and globally determines a wave speed reduction. 
The concept can be generalized to multi-dimensional or multi-physics systems characterized by a coupling induced between waves of different nature, such as in piezo-phononic structures \cite{Bergamini2015,Alan2019}, in micro electromechanical systems, or through fluid-structure interactions \cite{Colquitt2018}. This may open opportunities in the context of wave manipulation and control in metamaterial structures with concurrent wave conversion and trapping capabilities. Concerning future developments in this direction, our implementation can be easily scaled and adapted to operate at the microscale, to pursue energy harvesting and wave conversion.

\section*{Acknowledgements}
The support of the H2020 FET-proactive project MetaVEH under grant agreement No. 952039 is acknowledged. We also gratefully acknowledge the Italian Ministry of Education, University and Research for the support provided through the Project “Department of Excellence LIS4.0—Lightweight and Smart Structures for Industry 4.0.”

%\tableofcontents

% \bibliographystyle{elsarticle-num} 
% \bibliography{apssamp.bib}{}

% \bibliography{apssamp}% Produces the bibliography via BibTeX.

\end{document}

\preprint{APS/123-QED}

\title{Selective mode conversion and rainbow trapping via graded elastic waveguides: Supplementary Material}% Force line breaks with \\

\author{Jacopo M. De Ponti$^{1}$, Luca~Iorio$^{1}$, Emanuele Riva$^2$, Raffaele Ardito$^{1}$, Francesco Braghin$^2$, Alberto Corigliano$^1$}

\affiliation{$^1$ Dept. of Civil and Environmental Engineering, Politecnico di Milano, Piazza Leonardo da Vinci, 32, 20133 Milano, Italy \\
$^2$ Dept. of Mechanical Engineering, Politecnico di Milano, Via Giuseppe La Masa, 1, 20156 Milano, Italy\\ 
} 

% Calcoli modello analitico
% Sistema non graded
% Dati setup sperimentale
% Input Torsione e Input S0

\maketitle
\beginsupplement

To aid insight into the underlying physics of selective mode conversion and rainbow trapping, we give a detailed description of the analytical procedures, numerical models/additional analyses, sample manufacturing and experimental methods.

\section{SUPPLEMENTARY NOTE 1: NUMERICAL PROCEDURE FOR THE PLANE WAVE EXPANSION METHOD}
Consider the elastodynamic equation for an Euler Bernoulli beam subjected to flexural-torsional coupling:

\begin{equation}
\begin{split}
    EI_yw_{,xxxx}+\rho Aw_{,tt}&=
    {\sum_{r=1}^{2}\sum_{n=-\infty}^{+\infty}}m\omega_0^2\bigg[\Psi_r\left(x_r+na\right)+\\[4pt]
    &-w\left(x_r+na\right)+(-1)^r\frac{b+l}{2}\phi_r\left(x_r+na\right)\bigg]\delta\left(x-x_r-na\right)\\[6pt]
    GJ_t\phi_{,xx}-\rho J_p\phi_{,tt}&=-{\sum_{r=1}^{2}\sum_{n=-\infty}^{+\infty}}(-1)^rm\omega_0^2\frac{b+l}{2}\bigg[\Psi_r\left(x_r+na\right)+\\[4pt]
    &-w\left(x_r+na\right)+(-1)^r\frac{b+l}{2}\phi_r\left(x_r+na\right)\bigg]\delta\left(x-x_r-na\right)\\[6pt]
    \Psi_{r_,tt}\left(x_r+na\right)+\omega_0^2&\bigg[\left(-1\right)^r\frac{b+l}{2}\phi\left(x_r+na\right)-w\left(x_r+na\right)+\Psi_r\left(x_r+na\right)\bigg]=0
\end{split}
\label{eq:S01}
\end{equation}

Where $EI_y$, $\rho A$ are the flexural stiffness and the linear mass density of the beam, $I_y$ is the second moment of area $I_y=bh^3/12$. $b=7\;{\rm mm}$ and $h=2\;{\rm mm}$ are the width and the thickness of the beam. $E=70\;{\rm GPa}$ is the Young's Modulus, $\rho=2710\;{\rm Kg/m^3}$ is the material density and $A=bh$ is the cross-section area. $GJ_t$ and $\rho J_p$ are the rotary stiffness and inertia. $G=E/2(1+\nu)$ is the shear modulus ($\nu=0.33$), $J_t$ is the torsional constant, and $J_p$ is the polar second moment of area. The resonator stiffness and mass are $k=3EI_{res}/l^3$ and $m=\rho chl$, where $I_{res}=ch^3/12$ is the second moment of area of the resonator. $l$, $h=2\;{\rm mm}$ and $c=5\;{\rm mm}$ are the length, thickness, and width of the resonator. $\omega_0=\sqrt{k/m}$ is the bending resonance frequency of the attachments. \\
In the attempt to evaluate the dispersion properties $\omega\left(\kappa_x,l\right)$ of the undelying medium, Ansatz are considered for the displacement $w\left(x,t\right)=\hat{w}\left(x\right){\rm e}^{-j(\kappa_x x-\omega t)}$ and for the rotation $\phi\left(x,t\right)=\hat{\phi}\left(x\right){\rm e}^{-j(\kappa_x x-\omega t)}$ where $\hat{w}\left(x\right)=\sum_{p=-P}^{P}\hat{w}_p{\rm e}^{-jn\kappa_mx}$, $\hat{\phi}\left(x\right)=\sum_{p=-P}^{P}\hat{\phi}_p{\rm e}^{-jn\kappa_mx}$ reflect the periodicity of the medium and $\kappa_m=2\pi/a$ is the modulation wavenumber. Harmonic motion is assumed for the resonator displacement $\Psi_r\left(x_r\right)=\hat{\Psi}_r\left(x_r\right){\rm e}^{j\omega t}$.
Due to the periodic nature of the underlying medium, the relation between displacements and rotations between consecutive units writes:

\begin{equation}
w\left(x_r+na\right)=w\left(x_r\right){\rm e}^{-j\kappa na}\hspace{0.75cm}\phi\left(x_r+na\right)=\phi\left(x_r\right){\rm e}^{-j\kappa na}\hspace{0.75cm}\Psi_r\left(x_r+na\right)=\Psi_r\left(x_r\right){\rm e}^{-j\kappa na}
\label{eq:S02}
\end{equation}

plugging Equation \ref{eq:S02} into Equation \ref{eq:S01} gives:

\begin{equation}
\begin{split}
    EI_yw_{,xxxx}+\rho Aw_{,tt}&=
    {\sum_{r=1}^{2}\sum_{n=-\infty}^{+\infty}}m\omega_0^2\bigg[\Psi_r\left(x_r\right)+\bigg.\\[4pt]
    &\left.-w\left(x_r\right)+(-1)^r\frac{b+l}{2}\phi_r\left(x_r\right)\right]{\rm e}^{-j\kappa na}\delta\left(x-x_r-na\right)\\[6pt]
    GJ_t\phi_{,xx}-\rho J_p\phi_{,tt}&=-{\sum_{r=1}^{2}\sum_{n=-\infty}^{+\infty}}(-1)^rm\omega_0^2\frac{b+l}{2}\bigg[\Psi_r\left(x_r\right)+\bigg.\\[4pt]
    &\left.-w\left(x_r\right)+(-1)^r\frac{b+l}{2}\phi_r\left(x_r\right)\right]{\rm e}^{-j\kappa na}\delta\left(x-x_r-na\right)\\[6pt]
    \Psi_{r_,tt}\left(x_r\right)+\omega_0^2&\left[\left(-1\right)^r\frac{b+l}{2}\phi\left(x_r\right)-w\left(x_r\right)+\Psi_r\left(x_r\right)\right]=0
\end{split}
\label{eq:S03}
\end{equation}

The right hand side of Equation \ref{eq:S03} is non-null if $na=x-x_r$, which allows to eliminate the dependence on $n$ in the exponential term. Equation \ref{eq:S03} is conveniently rewritten as:

\begin{equation}
\begin{split}
    EI_yw_{,xxxx}+\rho Aw_{,tt}&=
    {\sum_{r=1}^{2}\sum_{n=-\infty}^{+\infty}}m\omega_0^2\bigg[\Psi_r\left(x_r\right)+\bigg.\\[4pt]
    &\left.-w\left(x_r\right)+(-1)^r\frac{b+l}{2}\phi_r\left(x_r\right)\right]{\rm e}^{-j\kappa\left(x-x_r\right)}\delta\left(x-x_r-na\right)\\[6pt]
    GJ_t\phi_{,xx}-\rho J_p\phi_{,tt}&=-{\sum_{r=1}^{2}\sum_{n=-\infty}^{+\infty}}(-1)^rm\omega_0^2\frac{b+l}{2}\bigg[\Psi_r\left(x_r\right)+\bigg.\\[4pt]
    &\left.-w\left(x_r\right)+(-1)^r\frac{b+l}{2}\phi_r\left(x_r\right)\right]{\rm e}^{-j\kappa\left(x-x_r\right)}\delta\left(x-x_r-na\right)\\[6pt]
    \Psi_{r_,tt}\left(x_r\right)+\omega_0^2&\left[\left(-1\right)^r\frac{b+l}{2}\phi\left(x_r\right)-w\left(x_r\right)+\Psi_r\left(x_r\right)\right]=0
\end{split}
\label{eq:S04}
\end{equation}

The term $\sum_{n=-\infty}^{\infty}\delta\left(x-x_r-na\right)=1/a\sum_{n=-\infty}^{\infty}{\rm e}^{jn\kappa_m\left(x-x_r\right)}$ can be be written in terms of exponential functions, where $1/a$ are the Fourier coefficients of the expansion. We are now ready to enforce the Bloch-wave solution for the displacements and rotations. We get to:

\begin{equation}
\resizebox{.99\hsize}{!}{$
\begin{aligned}
    EI_y\sum_{p=-\infty}^{\infty}&\left(\kappa+p\kappa_m\right)^4\hat{w}_p{\rm e}^{-j\left(\kappa+p\kappa_m\right)x}-\omega^2\rho A\sum_{p=-\infty}^{\infty}\hat{w}_p{\rm e}^{-j\left(\kappa+p\kappa_m\right)x}=
    \frac{m\omega_0^2}{a}{\sum_{r=1}^{2}\sum_{n=-\infty}^{+\infty}}\Bigg[\Psi_r{\rm e}^{jn\kappa_m\left(x-x_r\right)}+\Bigg.\\[4pt]
    -&\sum_{p=-\infty}^{\infty}\hat{w}_p{\rm e}^{-j\left(\kappa+p\kappa_m\right)x_r}{\rm e}^{jn\kappa_m\left(x-x_r\right)}+\left.(-1)^r\frac{b+l}{2}\sum_{p=-\infty}^{\infty}\hat{\phi}_p{\rm e}^{-j\left(\kappa+p\kappa_m\right)x_r}{\rm e}^{jn\kappa_m\left(x-x_r\right)}\right]{\rm e}^{-j\kappa\left(x-x_r\right)}\\[6pt]
    GJ_t\sum_{p=-\infty}^{\infty}&\left(\kappa+p\kappa_m\right)^2\hat{\phi}_p{\rm e}^{-j\left(\kappa+p\kappa_m\right)x}-\rho J_p\omega^2\sum_{p=-\infty}^{\infty}\hat{\phi}_p{\rm e}^{-j\left(\kappa+p\kappa_m\right)x}=-
    \frac{m\omega_0^2}{a}\frac{b+l}{2}{\sum_{r=1}^{2}(-1)^r\sum_{n=-\infty}^{+\infty}}\Bigg[\Psi_r{\rm e}^{jn\kappa_m\left(x-x_r\right)}+\Bigg.\\[4pt]
    -&\sum_{p=-\infty}^{\infty}\hat{w}_p{\rm e}^{-j\left(\kappa+p\kappa_m\right)x_r}{\rm e}^{jn\kappa_m\left(x-x_r\right)}+\left.(-1)^r\frac{b+l}{2}\sum_{p=-\infty}^{\infty}\hat{\phi}_p{\rm e}^{-j\left(\kappa+p\kappa_m\right)x_r}{\rm e}^{jn\kappa_m\left(x-x_r\right)}\right]{\rm e}^{-j\kappa\left(x-x_r\right)}\\[6pt]
    -\omega^2\hat{\Psi}_{r}&+\omega_0^2\left[\left(-1\right)^r\frac{b+l}{2}\sum_{p=-\infty}^{\infty}\hat{\phi}_p{\rm e}^{j\left(\kappa+p\kappa_m\right)x}-\sum_{p=-\infty}^{\infty}\hat{w}_p{\rm e}^{j\left(\kappa+p\kappa_m\right)x}+\hat{\Psi}_r\right]=0
\end{aligned}$}
\label{eq:S05}    
\end{equation}

for ease of notation, the resonator displacement at a distance $x_r$ is replaced with $\Psi_r\left(x_r\right)=\Psi_r$. We now multiply by an orthogonal function ${\rm e}^{js\kappa_mx}$ and integrate over the unit cell spatial domain $D\in\left[0,a\right]$.\\
The equations are simplified as:

\begin{equation}
\resizebox{.99\hsize}{!}{$
\begin{aligned}
    EI_y\left(\kappa+s\kappa_m\right)^4\hat{w}_s-\omega^2\rho A\hat{w}_s-
    \frac{m\omega_0^2}{a}{\sum_{r=1}^{2}}\Bigg[\Psi_r{\rm e}^{j\left(\kappa+s\kappa_m\right)x_r}&-\Bigg.\left.\sum_{p=-\infty}^{\infty}\hat{w}_p{\rm e}^{-j\left[p-s\right]\kappa_mx_r}+(-1)^r\frac{b+l}{2}\sum_{p=-\infty}^{\infty}\hat{\phi}_p{\rm e}^{-j\left[p-s\right]\kappa_mx_r}\right]=0\\[7pt]
    GJ_t\left(\kappa+s\kappa_m\right)^2\hat{\phi}_s-\rho J_p\omega^2\hat{\phi}_s+
    \frac{m\omega_0^2}{a}\frac{b+l}{2}{\sum_{r=1}^{2}(-1)^r}\Bigg[\Psi_r&{\rm e}^{j\left(\kappa+s\kappa_m\right)x_r}-\sum_{p=-\infty}^{\infty}\hat{w}_p{\rm e}^{-j\left[p-s\right]\kappa_mx_r}+\\[4pt]
    &\hspace{1cm}\left.+(-1)^r\frac{b+l}{2}\sum_{p=-\infty}^{\infty}\hat{\phi}_p{\rm e}^{-j\left[p-s\right]\kappa_mx_r}\right]=0\\[7pt]
    -\omega^2\hat{\Psi}_{r}+\omega_0^2\Bigg[\left(-1\right)^r\frac{b+l}{2}\sum_{p=-\infty}^{\infty}\hat{\phi}_p{\rm e}^{j\left(\kappa+p\kappa_m\right)x}-\sum_{p=-\infty}^{\infty}\Bigg.\Bigg.\hat{w}_p&{\rm e}^{j\left(\kappa+p\kappa_m\right)x}+\hat{\Psi}_r\Bigg]=0
\end{aligned}
$}
\label{eq:S06}    
\end{equation}

Equation \ref{eq:S06} is truncated to the first $s\in\left[-P,P\right]$, $p\in\left[-P,P\right]$ plane wave components and is written in a matrix form, making explicit the equations for the consecutive resonators:

\begin{equation}
    \begin{cases}
    K_1\hat{\bm{w}}-\omega^2M_1\hat{\bm{w}}+\Lambda_{11}\hat{\Psi}_1+\Lambda_{12}\hat{\Psi}_2+C_{11}\hat{\bm{w}}+C_{12}\hat{\bm{\phi}}=0\\[5pt]
    K_2\hat{\bm{\phi}}-\omega^2M_2\hat{\bm{\phi}}+\Lambda_{21}\hat{\Psi}_1+\Lambda_{22}\hat{\Psi}_2+C_{21}\hat{\bm{w}}+C_{22}\hat{\bm{\phi}}=0\\[5pt]
    \Gamma_{11}\hat{\bm{w}}+\Gamma_{12}\hat{\bm{\phi}}+\omega_0^2\hat{\Psi}_1-\omega^2\hat{\Psi}_1=0\\[5pt]
    \Gamma_{21}\hat{\bm{w}}+\Gamma_{22}\hat{\bm{\phi}}+\omega_0^2\hat{\Psi}_2-\omega^2\hat{\Psi}_2=0\\[5pt]
    \end{cases}
\end{equation}

where the displacement and rotation vectors $\hat{\bm{w}}$ and $\hat{\bm{\phi}}$ accommodate the expansion coefficients $\hat{\bm{w}}=\left[w_{-P},\ldots,w_{P}\right]$, $\hat{\bm{\phi}}=\left[\phi_{-P},\ldots,\phi_{P}\right]$. The above matrices are defined as:

\begin{equation}
\begin{split}
&K_1=\begin{bmatrix}
\ddots& & \\
 &EI_y\left(\kappa+s\kappa_m\right)^4& \\
 & &\ddots
\end{bmatrix}\hspace{0.5cm}
K_2=\begin{bmatrix}
\ddots& & \\
 &GJ_t\left(\kappa+s\kappa_m\right)^2& \\
 & &\ddots
\end{bmatrix}\\[5pt]
&C_{11}=\frac{m\omega_0^2}{a}\begin{bmatrix}
 &\vdots & \\
\hdots&\sum_{r=1}^{2}{\rm e}^{j\left[s-p\right]\kappa_mx_1}&\hdots\\
 &\vdots & 
\end{bmatrix}\hspace{0.5cm}
C_{12}=-\frac{m\omega_0^2}{a}\frac{b+l}{2}\begin{bmatrix}
 &\vdots & \\
\hdots&\sum_{r=1}^{2}(-1)^r{\rm e}^{j\left[s-p\right]\kappa_mx_1}&\hdots\\
 &\vdots & 
\end{bmatrix}\\[6pt]
&C_{21}=C_{12}\hspace{0.5cm}
C_{22}=\frac{m\omega_0^2}{a}\frac{b+l}{2}\begin{bmatrix}
 &\vdots & \\
\hdots&\sum_{r=1}^{2}{\rm e}^{j\left[s-p\right]\kappa_mx_1}&\hdots\\
 &\vdots & 
\end{bmatrix}\\[6pt]
&\Lambda_{11}=-\frac{m\omega_0^2}{a}\begin{bmatrix}
\vdots\\
{\rm e}^{j\left(\kappa+s\kappa_m\right)x_1}\\
\vdots
\end{bmatrix}\hspace{0.5cm}
\Lambda_{12}=-\frac{m\omega_0^2}{a}\begin{bmatrix}
\vdots\\
{\rm e}^{j\left(\kappa+s\kappa_m\right)x_2}\\
\vdots
\end{bmatrix}\\[6pt]
&\Lambda_{21}=-\frac{m\omega_0^2}{a}\frac{b+l}{2}\begin{bmatrix}
\vdots\\
{\rm e}^{j\left(\kappa+s\kappa_m\right)x_1}\\
\vdots
\end{bmatrix}\hspace{0.5cm}
\Lambda_{22}=\frac{m\omega_0^2}{a}\frac{b+l}{2}\begin{bmatrix}
\vdots\\
{\rm e}^{j\left(\kappa+s\kappa_m\right)x_2}\\
\vdots
\end{bmatrix}\\[6pt]
&\Gamma_{11}=-\omega_0^2\begin{bmatrix}
\hdots&{\rm e}^{-j\left(\kappa+s\kappa_m\right)x_1}&\hdots
\end{bmatrix}\hspace{0.5cm}
\Gamma_{12}=-\omega_0^2\frac{b+l}{2}\begin{bmatrix}
\hdots&{\rm e}^{-j\left(\kappa+s\kappa_m\right)x_1}&\hdots
\end{bmatrix}\\[6pt]
&\Gamma_{21}=-\omega_0^2\begin{bmatrix}
\hdots&{\rm e}^{-j\left(\kappa+s\kappa_m\right)x_2}&\hdots
\end{bmatrix}\hspace{0.5cm}
\Gamma_{22}=\omega_0^2\frac{b+l}{2}\begin{bmatrix}
\hdots&{\rm e}^{-j\left(\kappa+s\kappa_m\right)x_2}&\hdots
\end{bmatrix}
\end{split}
\end{equation}

and $M_1=\rho AI$ and $M_2=\rho J_pI$ are the mass and inertia matrices. $I$ is the $\left(2P+1\right)\times\left(2P+1\right)$ identity matrix. 
The vector coefficients $\hat{\bm{w}}$, $\hat{\bm{\phi}}$ and the resonator displacements $\Psi_{1,2}$ are stored in $\bm{\hat{\eta}}=\left[\bm{\hat{w}},\bm{\hat{\phi}},\hat{\Psi}_1,\hat{\Psi}_2\right]^T$ for a more compact representation of the dynamic equations, yielding the dispersion relation:

\begin{equation}
    K\left(\kappa_x,l_r\right)\bm{\hat{\eta}}=\omega^2M\bm{\hat{\eta}}
\end{equation}

where $K$ and $M$ are the $(2P+2)\times (2P+2)$ wavenumber dependent stiffness and mass matrices of the unit cell.

\section{SUPPLEMENTARY NOTE 2: Numerical Methods}

The numerical models employed for the transient analysis are  based on the commercial software Abaqus and, specifically, a finite element discretisation is accomplished through full 3D stress quadratic elements (C3D20). The analysis is performed via implicit time integration based on the Hilber-Hughes-Taylor operator \cite{Implicit}, an extension of the Newmark $\beta$-method with a constant time increment $dt= 0.02$ ${\rm ms}$. The system is forced through an imposed pressure $p(t)$ applied to a rectangular surface $S_0 = 42.4$ ${\rm mm^2}$. The input signal is a finite burst $p(t) = p_0\sin{(2\pi f_c)}{\rm w}(n)$ with amplitude $p_0 = 1\;{\rm MPa}$, where ${\rm w}(n)$ is a Hann window with time duration $15$ ${\rm ms}$. As a result, the input signal is charaterized by central frequency $f_c = 2.12$ ${\rm kHz}$ and a spectral content of width $\Delta f = 0.14$ ${\rm kHz}$.\\
In the numerical part of the paper, an infinite waveguide model is employed to avoid spurious edge reflections. This is achieved by imposing absorbing boundary conditions at both ends of the strip. The absorbing boundaries are implemented in Abaqus using the ALID (Absorbing Layers using Increasing Damping) method, as widely used in the elastic wave community, adopting a cubic law function for mass proportional Rayleigh damping with ${c}_{M\max }={10}^{6}$. A detailed description is provided in \cite{RAJAGOPAL201230}. The absorbing boundary is discretized using $100$ finite elements along $x$, covering a total length of approximately $3\lambda_{max}$, where $\lambda_{max}$ is the maximum wavelength travelling in the system.

\section{SUPPLEMENTARY NOTE 3: Additional numerical analyses}
\begin{figure*}[h]
\centering
\subfloat[]{\includegraphics[width=0.47\textwidth]{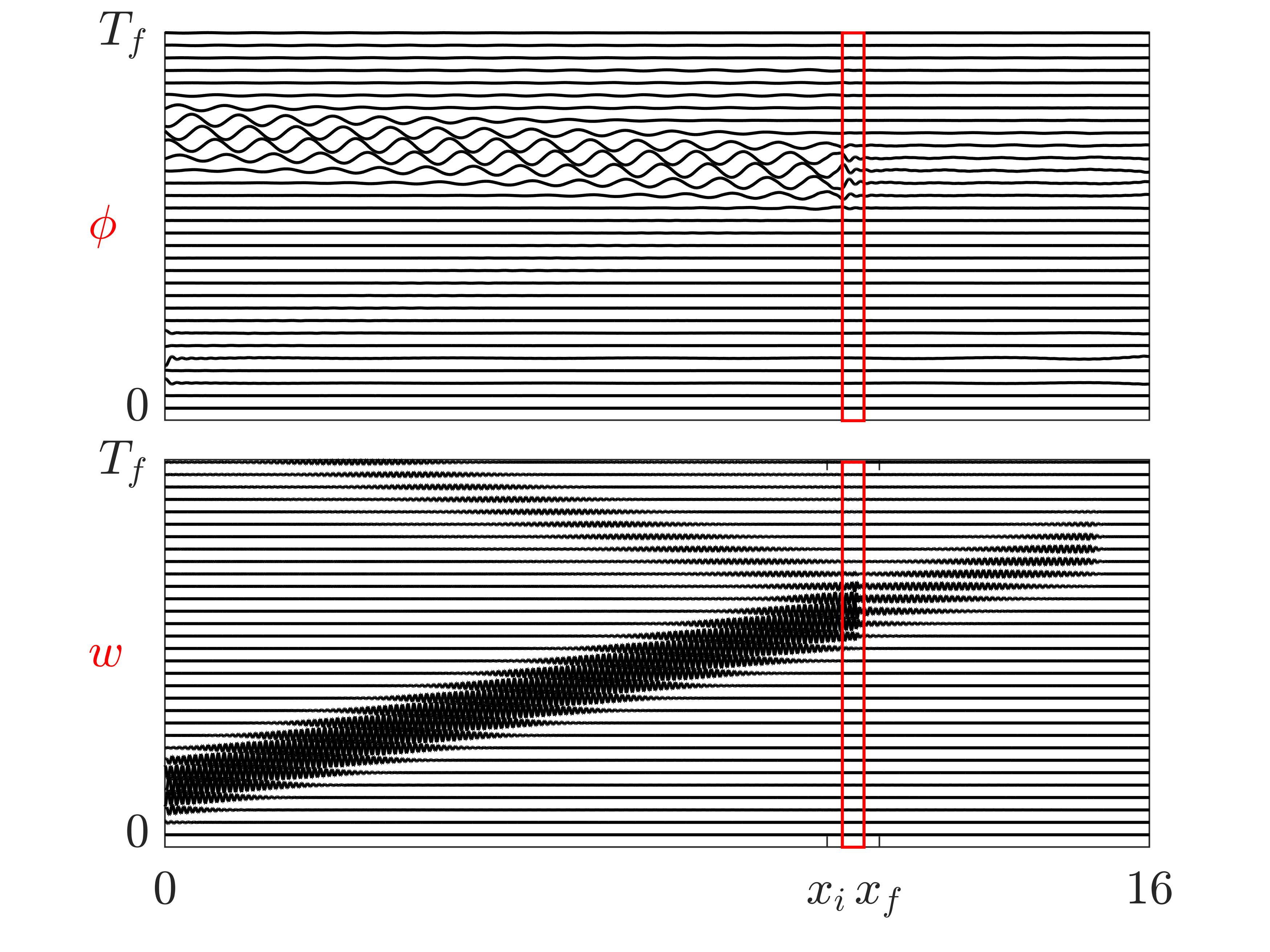}}
\subfloat[]{\includegraphics[width=0.47\textwidth]{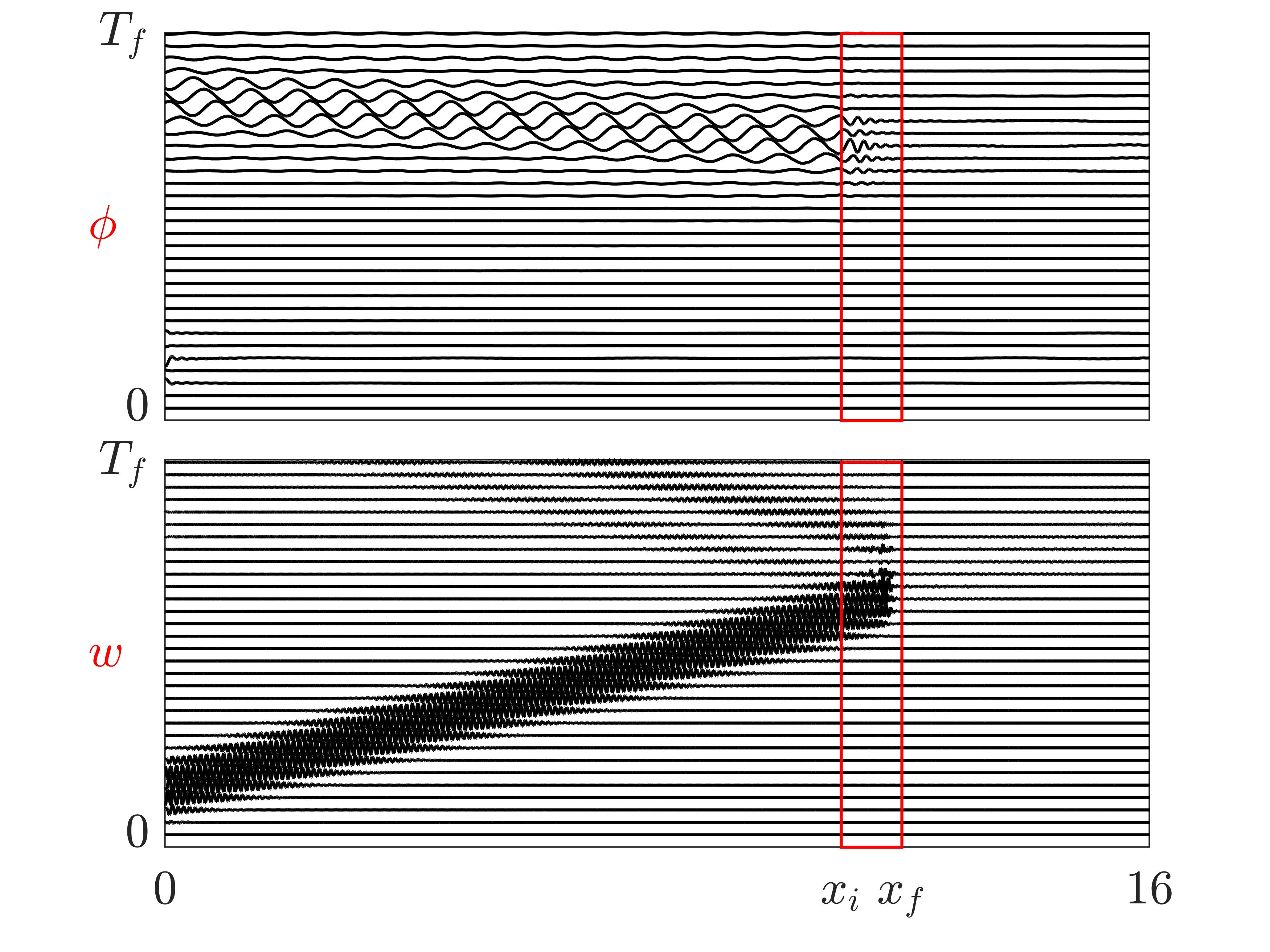}}\\
\subfloat[]{\includegraphics[width=0.47\textwidth]{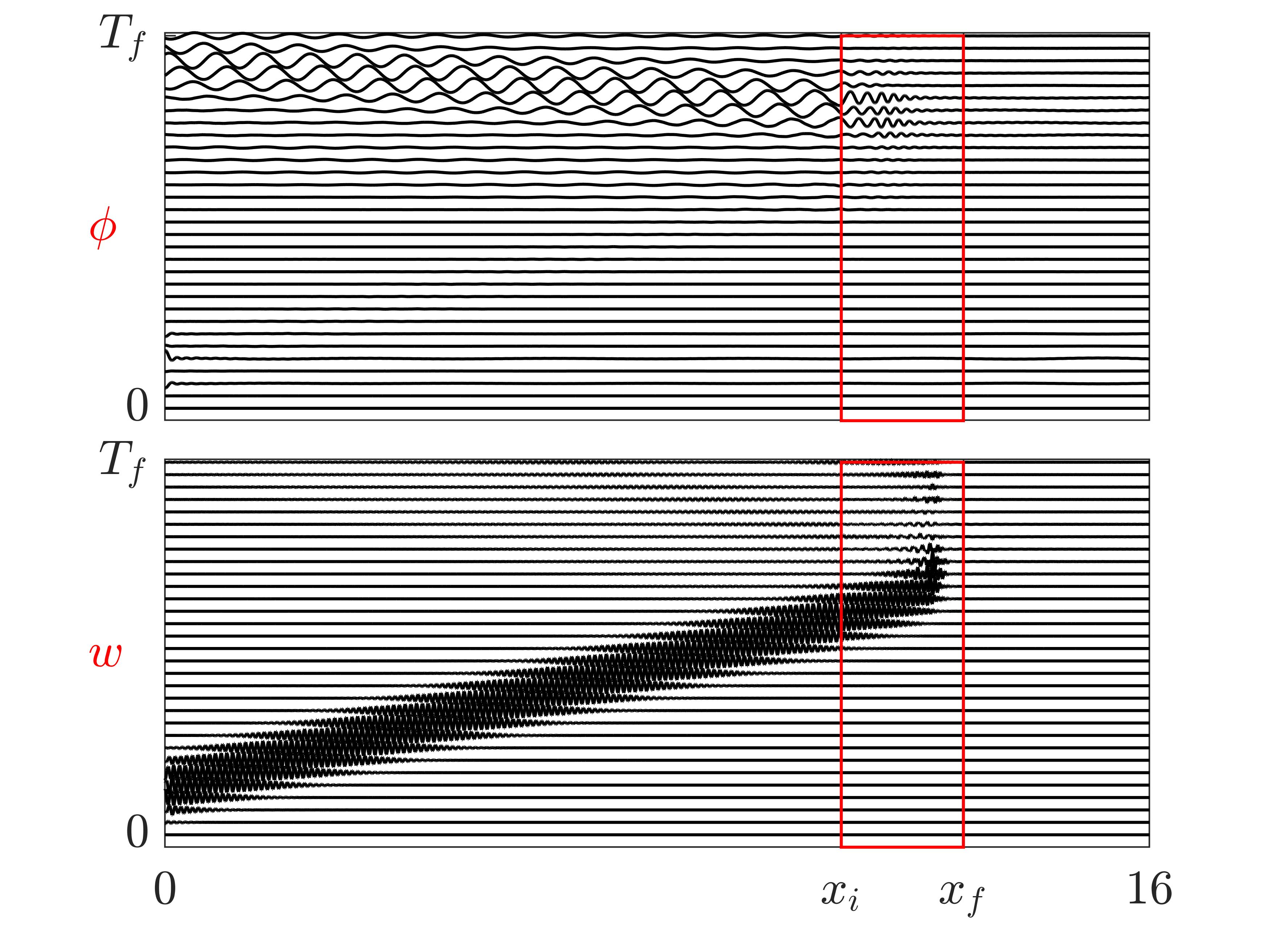}}
\caption{Space-time waterfall plots illustrating the vertical displacement $w$ (bottom) and the rotation $\phi$ (top) along the waveguide. The array of elements is highlighted with a red box. The numerical simulations are performed for a configuration with an array of (a) $9$, (b) $25$, and (c) $50$ pair of resonators. The torsional and flexural waves represented in the diagrams are normalized by the same amplitude coefficient.}
\label{fig:SW}
\end{figure*}

In this supplementary note, we present a number of additional analyses in the attempt to shed light on the conversion efficiency between propagating waves, as well as to corroborate the underlying physics illustrated through Fig. 2(d) in the main text. To this end, we have performed three distinct transient simulations upon varying the number of symmetry-broken pair of resonators ($9$, $25$, and $50$) and keeping unaltered the length of the first and last resonator of the array. The resonators are placed on a longer beam, which is characterized by an initial part of $13$ m. This allows a better visualization of the propagating wave packet. Also, the left and right ends are equipped with absorbing boundary conditions. According with the theoretical arguments, which are detailed in the main text, for a system made of an infinite number of resonators, the wave propagation follows a wavenumber transformation that brings the group velocity to zero. In contrast, if the system is finite, the incident flexural wave is scattered as a negative propagating component and wavenumber transformed from a wave with mixed polarization to a purely torsional wave, provided that the conversion efficiency is close to $100\%$. This is the expected behavior in the cases under analysis. The numerical results are shown in Fig. \ref{fig:SW}(a-c).

In case of $9$ pair of resonators (Fig. \ref{fig:SW}(a)), most of the energy is converted into a back-traveling torsional wave, while only a small amount of energy leaks throughout the array or is back-reflected as a flexural wave. Such a reflection or transmission is less visible in case of $25$ and $50$ pair of resonators, as shown in Fig. \ref{fig:SW}. This demonstrates that, due to the physical principle illustrated in Fig. 2(d) and for a sufficiently long array of elements, the conversion efficiency tends to $100\%$. As a final remark, it is worth to mention that for a longer implementation of the array, the slow-down process and the wavenumber transformation is more visible, which further confirms the theoretical arguments detailed in the main text.

\section{Supplementary note 4: Sample manufacturing and experimental methods}

\begin{figure}[h]
\centering
\includegraphics[width=0.95\textwidth]{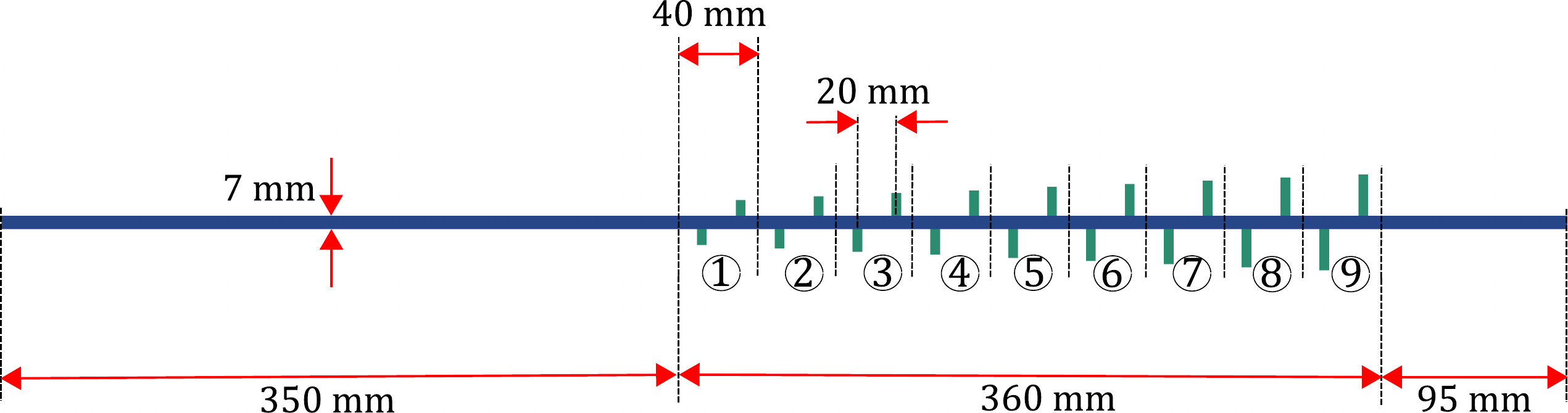}
\caption{Top view of the system, made of a homogeneous beam rigidly connected to a graded array of resonators.}
\label{fig:01}
\end{figure}

The system is manufactured from a $2$ ${\rm mm}$ thick aluminium plate through a laser cutting technology. A Bystronic Fibra (3000 W) laser machine is used to cut the slab with a position tolerance of $\pm 0.1\;{\rm mm}$.
A top view of the system is shown in Fig. \ref{fig:01}, which is made of an elastic waveguide $850$ ${\rm mm}$ long, $7$ ${\rm mm}$ wide and $2$ ${\rm mm}$ thick. The system includes an initial portion $350\;{\rm mm}$ long that serves as an input domain.
The final shape of the specimen includes the array of resonators, thereby avoiding additional operations such as soldering or mechanical connections that would have led to imperfections in the prototype. The array is composed of $9$ unit cell of size $a=40$ ${\rm mm}$, each one containing two identical resonators shifted by a distance $\xi = 20$ ${\rm mm}$. The geometrical dimensions of the resonators for each unit cell are stored in Tab. \ref{TableGeo} and numbered from 1 to 9. 

\begin{table}[h]
\begin{tabular}{rlrrrrrrrrr}
\hline
CELL NUMBER: &\hspace{1cm} \large \ding{172} & \hspace{1cm} \large \ding{173} &\hspace{1cm} \large \ding{174} &\hspace{1cm} \large \ding{175} &\hspace{1cm} \large \ding{176} &\hspace{1cm} \large \ding{177} &\hspace{1cm} \large \ding{178} &\hspace{1cm} \large \ding{179} &\hspace{1cm} \large \ding{180} & \\
\hline
$l_i$ (mm): &\hspace{1cm} 8.1 & 10.0 & 11.6 & 13.0 & 14.8 & 16.3 & 18.0 & 19.7 & 21.2\\
\hline
$c_i$ (mm): &\hspace{1cm} 5 \\
\hline
$h_i$ (mm): &\hspace{1cm} 2 \\
\hline
\end{tabular}
\caption{Geometric parameters of the graded array of resonators. Each unit cell contains two identical resonators of length $l_i$, width $c_i$, and thickness $h_i$, shifted by a constant distance of $\xi = 20$ ${\rm mm}$.}
\label{TableGeo}
\end{table}

The experimental tests are performed according to the following setup. The waveguide is mechanically joined to a LDS v406 electrodynamic shaker at the left boundary, to provide an out-of-plane input excitation and avoid the excitation of torsional and longitudinal wave modes. The opposite end is clamped, to avoid undesired deformations of the beam. 
A reflective powder is employed to cover the beam, to increase the quality of the measurements, which are performed through a Polytec 3D scanner laser Doppler vibrometer (SLDV), that is able to measure and separate the 3D velocity field $u(x,y)$, $v(x,y)$, and $w(x,y)$ on the top surface of the beam. Finally, the excitation signal, provided through a KEYSIGHT 33500B waveform generator, synchronously starts with the measurement system and consists in a input function
$V(t) = V_0w(n)sin(2\pi f_c)$ with amplitude  $V_0= 2$ ${\rm V}$, Hann window $w(n)$, central frequency $f_c= 2.12$ ${\rm kHz}$ and time duration $15$ ${\rm ms}$.